# Set-theoretic solution for the tuning problem

by Vsevolod Deriushkin
for the newtonality.net

## Table of Contents

In this paper, I want to suggest a new solution to the problem of musical tuning. On the one hand, I see it as a generalization of Just Intonation (JI) to inharmonic timbers; on the other, as a unification of spectral interference and harmonicity contributions to consonance within a single framework. The main achievement of the work is the ability to quantify the phenomenon of musical consonance using set theory. That quantification is done by defining two measures of consonance: affinity and harmonicity. These measures naturally generate sets of intervals that can be used as dynamic tuning systems. The paper is aimed at a broad audience of people who may not be skilled in music, tuning theory, or mathematics. Thus, I will try to provide as many details and explanations as possible while keeping the paper concise.

# 1 Introduction

This chapter is aimed at the reader who is not familiar with tuning theory. Thus, if you know what a musical interval is, why it can be represented by a ratio such as 3/2, and why it is difficult to tune a guitar by ear, you can safely skip that chapter. For everybody else, let me briefly introduce the main concepts of tuning theory and explain why it is not a trivial task to tune a musical instrument.

## 1.1 Basic notions of tuning theory

The most basic notion in music is the notion of a **pitch**. All sounds are perceived as ranging from low-pitched to high-pitched. An example of a low-pitched sound is a male baritone voice, and of a high-pitched sound, a female soprano voice. The pitch is an absolute position of sound in that range and is directly linked to the frequency[1] with which the sounding body vibrates. Low pitches are produced with low-frequency vibrations, and high pitches with high-frequency vibrations. The range of all possible pitches is continuous, and in theory, there is an infinite number of pitches.

Since ancient times, people have noticed that when two pitches sound together, depending on their relative positions, they may sound musical or unmusical. The relative position of pitches is called an **interval** and can be expressed as a ratio between the frequencies of two pitches. If two pitches are the same (have the same frequency), the interval is described by the ratio 1/1 and is called a **unison**. If one of the pitches is twice as high as the other (has double the frequency), the interval will be described by the ratio 2/1 or 1/2, depending on the order of the pitches. For historical reasons, the lower pitch is considered the base pitch from which the interval is built. So, the interval 2/1 is called an **octave** and is built up from the base pitch. The interval 1/2 is also an octave, but that is built down from the base pitch. Octaves play a unique role in music, as pitches an octave apart are considered equivalent (the same note). Other prominent intervals are the **perfect fifth** 3/2, and the **perfect fourth** 4/3, together with the octave and unison; they are the most musical and smooth-sounding intervals.

[1] A detailed definition of frequency is given in the chapter 3.1

Given those four intervals, we can start combining them to get new intervals. The combination is performed by multiplying the corresponding interval ratios. For example, to get the interval of a perfect twelfth, we need to multiply the ratios of an octave and a perfect fifth $\frac{2}{1} \cdot \frac{3}{2} = \frac{6}{2} = \frac{3}{1}$. We can also combine intervals by going in the downward direction by using an inverted ratio. Thus, the interval that is attained by going an octave up and a perfect fifth down is $\frac{2}{1} \cdot \frac{2}{3} = \frac{4}{3}$ a perfect fourth!

Describing intervals as ratios is a classical method that goes back to Pythagoras. Since then, tuning theory has undergone many changes, and nowadays we use a tuning system called **12 tone equal temperament (12-tet)**, in which intervals are measured differently. In 12-tet, the octave is divided into 12 equal intervals called **semitones,** and each semitone consists of 100 **cents**.[2] All other intervals in 12-tet are made from the whole number of semitones. Perfect fourth is five semitones or 500 cents, perfect fifth is seven semitones, or 700 cents, and octave is twelve semitones or 1200 cents. Despite having the same names, 12-tet intervals are different from those defined as simple ratios. For example, the perfect fourth defined as 4/3 is approximately 498.045 cents, rather than 500 as in 12-tet. That 2-cent difference is very small, and those intervals will sound almost identical. Still, if we take the major third as 5/4, it corresponds to 386.3137 cents rather than 400, which does sound noticeably different. The only interval that is the same is the octave, as 1200 cents gives exactly the ratio of 2/1.

## 1.2 The tuning problem

Given that some intervals are musical and others are unmusical, it would be very beneficial for composers and instrument builders to have a defined system of musical intervals. Such a system is a **musical tuning**. One may notice that defining intervals as musical and unmusical contains a significant degree of subjectiveness. That subjectiveness is one of the roots of the tuning problem.

[2] The semitone of 12-tet can be represented as a number $\sqrt[12]{2}$ which is irrational, meaning it is impossible to represent it as ratio n/m.

By what standard should we deem an interval as musical or not? If the musicality of an interval arises solely from perception, then differences in perception between people will lead to different tuning systems. Perhaps, there may be a natural tuning system that is musical regardless of subjective judgment. If so, what is such a system? Those are the questions that baffled theoreticians for thousands of years and that are discussed in detail in Chapter 2. Different answers to those questions lead to different tuning theories. Even within a single European tradition, several tuning systems coexisted, with some gaining and others losing dominant status over time. Nowadays, the Western tuning landscape is monopolized by 12-tet. That monopoly is not absolute, as small communities of enthusiasts preserve traditional tuning systems and experiment with new ones.

Historically, there have been attempts to find a natural tuning system free of subjectivity, but the solutions have always faced difficulties in practical implementation. To get a feeling for why, let's see what problem an aspiring guitar player will face when trying to tune a regular 6-string guitar by ear. A guitar has six strings; the first string has the highest pitch, and the sixth string has the lowest pitch. The standard tuning has the interval of a major third between the second and third strings and the interval of a perfect fourth between all other pairs of neighboring strings (Figure 1.1).

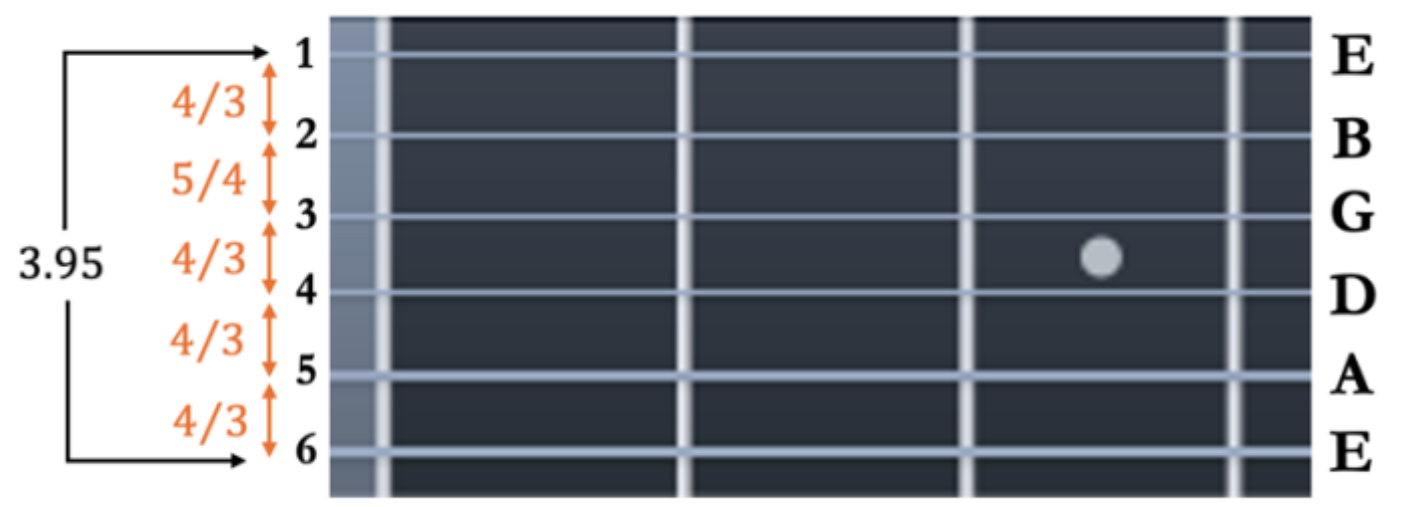

Figure 1.1 - Demonstration of the problem of tuning of 6-string guitar to just ratios. Adding up the intervals between neighboring strings (in orange) creates the dissonant interval of approx. 3.95 instead of 4/1 between low and high E.

When tuning all those intervals by ear, we naturally pick out a 4/3 ratio for the perfect fourth and a 5/4 ratio for the major third. Even minor deviations from those exact ratios are perceivable for the careful listener. In the properly tuned guitar, the first and sixth strings should be exactly two octaves apart with a ratio of 4/1 and thus form a perfect consonance. However, that is not the case with the way we tuned the guitar, and the reason is purely arithmetic. The ratio between the first and the sixth strings is the product of the multiplication of intervals between all other neighboring strings, or $\frac{4}{3} \cdot \frac{4}{3} \cdot \frac{4}{3} \cdot \frac{5}{4} \cdot \frac{4}{3} = \frac{1280}{324} \approx 3.95$, which is a little bit short of 4/1, and sounds noticeably out of tune. Moreover, almost every chord on such a guitar will sound a little off, rendering the instrument unplayable. The solution to that problem is not straightforward, as it depends on our values, goals, and the subjectivity of our perception.

That is the heart of the tuning problem. The natural intervals to which our senses gravitate are challenging to use in musical instruments, and the subjective factor in the perception of musicality of intervals further complicates addressing this issue. In the following chapters, I will present my solution to that problem. Along the way, we will delve into philosophy, mathematics, physics, and psychoacoustics. The long, complicated journey will yield a mathematical description of consonance and a candidate for a natural tuning system.

# 2 History and methodology of tuning theory

In this chapter, I want to share my perspective on the historical and philosophical context surrounding the tuning problem. I aim to briefly outline the historical development of the main thought traditions within that field. Doing so helps us see a larger picture in which the modern tuning system is just one of many ways to solve the tuning problem. Moreover, we notice that the technological limitations that led to 12-tet are no longer relevant in the 21st century. Therefore, it might be time to explore other schools of thought and their impact on music and tuning. My analysis focuses on European musical traditions originating from Ancient Greece and does not include music from other civilizations. The purpose of this chapter is to explain my motivation for this work and to outline the philosophical and methodological foundations behind it.

## 2.1 Pythagoras, Ptolemy, Aristoxenus, and Early Modernity

The word consonant originates from Latin consonantia, meaning "harmony, agreement." It is directly related to the musicality of intervals discussed in Chapter 1.2 and describes an observation that some intervals sound smooth, stable, and harmonious, as if they fit together forming one larger whole. The opposite of consonance is dissonance, a feeling of pitches clashing, not “agreeing” with each other, sounding rough and unstable. When exploring musical tuning theory, it becomes clear that consonance is a core phenomenon that underlies different tuning systems. Therefore, the most important factor influencing the development of tuning theory is our understanding of the nature of consonance. For thousands of years, people have recognized that consonance can be judged by consulting our senses or by using reason.

The earliest formulation of a musical tuning system in Europe was by Pythagoras in Ancient Greece (6th century B.C.). There is no direct text that was left by Pythagoras himself, so his teaching is known to us by the works of his followers, called Pythagoreans. They believed that consciousness precedes matter and that material phenomena are manifestations of timeless Ideas. The famous saying “All things are Number”, which is attributed to Pythagoras, means that all material phenomena are measurable and can be represented as numbers. In that sense, musical pitch can be expressed as how often the musical body beats the air per unit of time, or what we now call a frequency. Consequently, a musical interval, being the ratio between two pitches, can also be represented as a number. In a Pythagorean view, those numerical relationships are what really define consonance, and reasoning about them yields a theory of musical tuning.

Pythagoreans concluded that ratios 1/1, 2/1, 3/2, and 4/3 define perfect consonances of unison, octave, perfect fifth, and perfect fourth. Therefore, they were placed at the basis of their tuning theory.[3] The fact that small deviations from those ratios, such as 41/30 for a perfect fourth, also sound quite consonant to our ears was seen as a flaw of the auditory system that is incapable of perceiving such small distinctions. Pythagoreans compared judging consonance by ear alone to judging distance by eyesight alone.

The opposing view to judge consonance, which relied more on the senses than on reason, also emerged in Ancient Greece and was formulated by Aristoxenus (4th century B.C.). He was one of Aristotle's most talented students, who wrote more than 400 works on various topics. In his work "Elements of Harmonics", he opposes Pythagoreans and gives preference to the senses as the main judge of consonance.[4] If a combination of notes is perceived as consonant, it is consonant. The role of reason in such a case is to serve the senses by helping musical practitioners in reaching their goals. Instead of using numeric ratios to represent intervals, Aristoxenus suggested measuring musical intervals as a sum of discrete quantities. That approach is very similar to the modern approach of measuring intervals in cents.

In the 2nd century A.D., in his treatise Harmonics, Ptolemy criticizes the approaches of Pythagoreans and Aristoxeneans and proposes his own view of the roles of reasoning and senses in judging consonance. Ptolemy shares the Pythagorean idealistic view that mind precedes matter but argues that the senses also play a large role, even given their limitations. He argued that it is unreasonable for some Pythagoreans to dismiss senses when they profoundly disagree with their theory. When the contradiction with theoretical prediction is undeniable, the theory must be reconsidered. In his view, true theory should be in harmony with both reason and senses.[5]

These are the three main schools of thought in tuning theory that I believe remain relevant to this day. They can be put into two categories. On one side, there are **idealistic** theories by Pythagoreans and Ptolemy that give precedence to conciseness and the realm of Ideas. On the other side, there is the **empirical** theory of Aristoxenus, which emphasizes the role of senses and material phenomena. Both categories employ sensory observations to inform the mind about reality, but they differ in the type of reasoning used.

---

[3] [15] Book 1, chapter 1.5.

[4] [16] p. 196: "There is a certain marvelous order which belongs to the nature of harmony in general; in this order every instrument, to the best of its ability, participates under the direction of that faculty of sense-perception on which they, as well as everything else in music, finally depend."

[5] [15] Book 1, chapters 1.6 - 1.9.

Idealistic theories employ a combination of abductive and deductive reasoning, which leads to is a top-down approach. At first, top-level axioms or key Ideas of the theory are guessed using intuition (abductive reasoning), and then, the consequences of those ideas are obtained using deduction. Pythagoras guessed that all things are numbers and that numbers 1, 2, 3, and 4 are sacred[6], thus concluding the consonance of 1/1, 2/1, 3/2, and 4/3 as all possible ratios between those numbers (not counting inverse ratios). Knowing that, one can use deduction to find other intervals in tuning. Pythagoreans used the process of stacking and combining consonant intervals to get their tuning system.[7] Hard idealistic theories, like Pythagorean theory, downplay the role of empirical and sensory argumentation, while soft idealistic theories, like Ptolemaic theory, seek to find harmony between reason and observations. Thus, in soft idealistic theory, the deduction should be used to find testable predictions of initial axioms to validate the theory. It may come as a surprise, but most, if not all, great theories in physics are soft, idealistic theories. Einstein guessed the constancy of the speed of light, which was later confirmed by observations, not the other way around. Similar guesses were made by Newton, Maxwell, Heisenberg, and other great scientists in their famous theories.

Empirical theories, such as Aristoxenian theory, are built from the bottom up using inductive reasoning. Initially, observations are gathered to formulate low-level ideas and principles. These principles can then be further inductively generalized until sufficient explanatory power is achieved, and the theory is considered complete. Such theories are often used in the social sciences and can be very effective for achieving practical goals. However, gaining fundamental knowledge of the universe's structure seems to be accomplished through idealistic theories.

Historically, in the West, an idealistic view of consonance was dominant until early Modernity, when Renaissance thinkers turned to ancient texts to resurrect the lost music of antiquity. The 16th-century debates between Gioseffo Zarlino and Vincenzo Galilei (the father of Galileo Galilei) provide insight into the philosophical split that happened within tuning theory.[8] At that time, the most common tuning system was Pythagorean, in which intervals of thirds were not considered to be consonant[9]. However, the senses of everyone who attended church services and listened to chorales perceived them as consonant intervals. That was a problem, as it meant that chorales were not using intervals of Pythagorean tuning, but something else. Another problem was the application of Pythagorean tuning to fretted and keyboard instruments. Instruments tuned in such a way were not suitable for performing triadic harmony and sounded out of tune in many tonalities.

---

[6] [17] "Whole number in the view of Pythagoreans had a sacred meaning. Number 1 represented a unity and origin of all things, number 2 symbolized female principle, number 3 male principle and number 4 symbolized justice. Number 10 that is a sum of numbers 1, 2, 3, 4 was most perfect and meant unity that arises from multiplicity. All 5 numbers formed a tetractys the holy object that was worshiped by Pythagoreans."

[7] [8] Chapter 4. The full list of intervals of Pythagorean tuning: 1/1, 256/243, 9/8, 32/27, 81/64, 4/3, 729/512, 3/2, 128/81, 27/16, 16/9, 243/128

[8] For the detailed analysis of the Galilei-Zarlino dispute refer to [11]

[9] Pythagorean tuning is obtained by stacking perfect fifths on top of each other (raising 3/2 to the n-th power) and transposing the result back to be within the range of an octave (between 1/1 and 2/1). The interval of major third

Gioseffo Zarlino set out to overcome those challenges by formulating a new tuning theory, which earned him the status of one of the greatest musical theoreticians of his time. Zarlino worked within the established idealistic tradition of thought and turned to Ptolemy for the explanation of triadic consonance. He reasoned that Ptolemy's intense diatonic scale, also known as the Ptolemaic sequence[10], was the natural tuning that was a fundamental part of reality, independent of anyone's perceptions or point of view. The major third in that tuning was represented by a much simpler ratio of 5/4 than the Pythagorean 81/64 and sounded much smoother. Zarlino argued that natural instruments, such as the human voice, naturally gravitate towards the Ptolemaic sequence when singing polyphonic music. Artificial instruments, like keyboard or fretted instruments, have fixed positions for notes and thus are incapable of producing the correct Ptolemaic Sequence in all tonalities. Therefore, as a sacrifice to make those instruments sound better, we must artificially temper the intervals we tune them to, using a reasonable theoretical method.[11]

Galilei was a student of Zarlino, but being a lute builder, he was much more practically oriented than his teacher. Combining that with a counter-cultural stance, it sent Galilei on a very different path in tackling the same theoretical issues. Galilei wrote several responses to Zarlino's works, arguing that there is no such thing as natural tuning and that all tuning systems are constructs of theoreticians[12]. He also argued that it is not correct to divide instruments into natural and artificial, choirs and instruments that accompany them are fundamentally equal and perform in the same obtained in such way is 81/64 and minor third is 32/27, which do not sound very pure to the ear. The Pythagorean tuning for the major scale consists of the following intervals: 1/1, 9/8, 81/64, 4/3, 3/2, 27/16, 243/128, 2/1.

[10] Ptolemy's intense diatonic scale also known as Ptolemaic sequence or 5-limit just major scale, consists of the following intervals: 1/1, 9/8, 5/4, 4/3, 3/2, 5/3. 15/8, 2/1. Note that it shares a lot of intervals with Pythagorean major scale, but ratios for major third, major sixth and major seventh are much simpler.

[11] In Ptolemaic sequence the intervals are not evenly spaced within the octave, so performing the same melody or a chord built from different notes will contain different intervals. Tempering shifts intervals from their just values in order to make a tuning more even. From idealistic perspective, tempering artificially redistributes dissonance across tuning. Very dissonant intervals become less dissonant at the price of consonant intervals becoming less consonant, thus this process should be done carefully, guided by a reasonable theory.

[12] [11] p.5: "Galilei demands that we question the assumed 'naturalness' of musical scales or tuning systems. In his view, there was no justification to conclude that Ptolemy's Syntonic tetrachord was any more natural than those attributed to Pythagoras, Archytas, or Didymus: all were human constructions and thus were 'artificial.'"

tempered tuning. He referred to Aristoxenus as an authority to argue a primacy of senses in music and the necessity of tempering intervals to the values established by sensory observations. [13]

Both Zarlino and Galilei argued in favor of tempering intervals, but from different perspectives. For Zarlino, it was a sacrifice to make artificial instruments playable, and that tempering should be theoretically justified. For Galilei, tempering was an obvious solution to the practical problem, and the trial-and-error experimentation could be used to determine the correct tuning. With time, the design of keyboard instruments was perfected. The great convenience that was provided by keyboard instruments made them a must-have instrument for any serious music practitioner or composer. Early tempered tunings were made to minimize tempering and to keep as many thirds pure as possible, adhering to Zarlino's worldview. At the same time, the entirety of music theory, notation, and tuning started to form a self-contained system of music that standardized a musical language, education, and composition. To perfect the whole system of music, the tuning had to be further tempered.[14,15] In the modern 12-tet system, all intervals apart from the octave are tempered, which, from an idealistic perspective, means they are dissonant.

## 2.2 Modern consonance research

Nowadays, tuning theory is rarely taught in music schools, and 12-tet intervals are presented to students as a fact of music without explanation. As a result of the gradual shift towards empiricism and materialism that has happened since the 16th century, the consonance of intervals is explained by appeal to the senses. If a musician desires to delve into the reasons why intervals are the way they are, he or she will be presented with the evolutionary narrative. In this view, tuning theory first emerged as the Pythagorean system, later improved by various meantone temperaments, and in the early XX century reached its perfected state in the modern 12-tet system. The merit by which 12-tet is judged as a superior system is the ability to freely modulate between different tonalities and its suitability for fretted and keyboard instruments. Natural preference towards simple ratio intervals is downplayed in favor of convenience, to the point that tempering is not thought of as a sacrifice, but rather as a natural and necessary procedure to make music that sounds in-tune. That narrative provides an illusion of completeness and is very convincing because it aligns with a broader philosophical worldview and everyday experience. We all listen to 12-tet music that sounds just fine.

[13] [11] p. 82: "Galilei then justified the tempering of individual tones and semitones through the authority of Aristoxenus and explained its application through a practical demonstration on the keyboard."
[14] [12] Chapter 17 "Equal temperaments".
[15] In older tempered tunings such as meantone tuning the thirds were tried to be kept pure. That lead to different tonalities having different color or flavor, as their triads were slightly different. Some tonalities were outright unusable because of very dissonant triads. To remove that problem and allow for free modulation and transposition, a further tempering had to take place that sacrificed the purity of triads leaving only the octave untouched.

If a musician wants to understand the origin of consonance or what they perceive as consonance or dissonance, there is a vast scientific literature on that topic. Such research is conducted at the intersection of music theory, acoustics, biology, and psychology. The methodology is purely empirical, relying on biological facts about the auditory system and behavioral studies as sources of knowledge about consonance. A good review of the modern state of research can be found in [1]. In this article, the authors describe three predictors of consonance that can be used to categorize all tuning theories into three families. Those predictors are cultural familiarity, harmonicity/periodicity, and spectral interference.

Theories based on cultural familiarity suggest that there is no inherent phenomenon that defines consonance; instead, perception of consonance is entirely subjective. A culture's musical tradition creates a consensus among individuals, conditioning them to perceive familiar music as consonant and unfamiliar music as dissonant. This thesis relies on two observations. First, the set of consonant intervals varies significantly across different cultures. Second, every human has a natural preference for things that are familiar rather than alien. Consequently, the music you grow up with will sound more pleasant, and this sense of pleasantness is what we call consonance. Therefore, tuning is entirely arbitrary and not grounded in objective reality.

Periodicity and harmonicity theories propose that our auditory system evolved to recognize harmonic (periodic) sounds from the environment. Discerning such sounds is vital for our survival, as they are often produced by humans and animals. The defining feature of harmonic sounds is that they have a pitch. The easier it is for our auditory system to identify that pitch, the more consonant the sound is. Periodicity and harmonicity theories are less subjective than cultural familiarity theories because there is an objective mechanism through which waves of different pitches interact to produce a single pitch (more on that in Chapter 3.3). This mechanism explains the natural preference for simple ratio intervals and our ability to tolerate small deviations from those ratios. Because the origin of consonance is sensory, the exact value for consonance of an interval depends on the mechanism our auditory system uses to detect a virtual pitch. Currently, there is no consensus on what that mechanism is, but, in general, predictions from harmonicity/periodicity theories align with conventional music theory. However, there are some observations that this family of theories cannot explain. One important example is the tuning of inharmonic sounds.[16]

The last family of consonance theories is spectral interference theories. They rely on the fact that each real sound is made of elementary vibrations called partials.[17] When several sounds are present simultaneously, their partials interact in our inner ear. These interactions can create a sense of dissonance. An important example of spectral interference theory is laid out by William Sethares in his work "Tuning, Timbre, Spectrum, Scale". Sethares builds on Helmholtz's idea [2] that dissonance can be defined as a rough beating[18] between partials of different sounds. Consonance, in that sense, is the absence of such rough beating. Sethares uses empirical data to quantify dissonance between pairs of partials, allowing him to calculate dissonance for any sounds at any interval. He plots the distribution of dissonance against interval, which he calls the dissonance curve, and uses it to derive tuning systems. The predictions of his theory align well with common music theory and with tuning systems used for inharmonic instruments. Sethares demonstrates this by predicting the tuning for the Gamelan[19] orchestra from the spectrum of individual instruments. The most exciting part is that Sethares' theory has a great practical potential, as it provides a vision for new instruments with dynamic tuning and inharmonic timbres. It has its limitations, as it does not give a meaningful tuning for sounds with a small number of partials. But most importantly, it does not explain why inharmonic sounds may sound dissonant even when no rough beating is present.[20]

All three families of theories have solid reasoning, but notice that they are all empirical theories that attribute the origin of consonance to the human auditory system and perception. Therefore, we can conclude that the empirical and Aristoxenean worldview is currently dominant in tuning theory. Having said that, the idealistic and Pythagorean tradition is not gone. It is alive today within the microtonal community of musicians and is known as Just Intonation (JI). JI is an octave-based tuning system built from intervals represented by simple numeric ratios. It is commonly accepted within that theory that the degree of consonance of an interval is inversely proportional to the size of the numbers used to define that interval.[21] This way, the Ptolemaic third 5/4 is more consonant than the Pythagorean third 81/64, as smaller numbers are used in its ratio. Every rational number is to some degree consonant, and the historical development of tuning is seen as a slow process of recognizing that truth[22]. Modern 12-tet tuning relies on irrational numbers for intervals such as multiples of $\sqrt[12]{2}$, which are dissonant from the JI perspective. The fact that many modern listeners prefer the 12-tet third over the Pythagorean third can be explained as an imperfection of our hearing and perception.

---

16 Example of inharmonic sound is a sound of church bell. If the sound sounds "metallic" it is probably inharmonic. The issue with such sounds is that conventional consonant intervals, such as octave, may sound dissonant while deviations from those intervals may sound consonant, as was demonstrated by William Sethares in [4]

17 Partial is an elementary sine wave vibration. A complex sound is made of many partials with different frequencies. Mathematically, partial is a prominent sine wave component in Fourier transform representation of the wave.

18 Beating is the phenomena that is a product of interference of two waves with close frequencies. For explanation refer to [4] Chapter 3 "Sound on sound".

19 Gamelan is the name of the traditional Indonesian music that utilizes highly inharmonic metallophones. The tuning of Gamelan varies from orchestra to orchestra and significantly differs from common 12-tet. That difference is so huge that it contradicts all common European knowledge of consonance of intervals. Sethares theory is capable of explaining the tuning of Gamelan and the variation of that tuning from orchestra to orchestra.

20 Sethares involves cultural familiarity argument to explain why harmonic sounds are preferred to inharmonic sounds and that it takes time to get used to inharmonic harmonies. However, he also admits that ear tends to reset after some time to the preference for harmonic sounds once again.

## 2.3 The hierarchy of musical interfaces

The popularity of the 12-tet tuning system remained unshaken despite discoveries made by scientific consonance theories and critiques by proponents of JI. I believe that there are three reasons for that: network effects, heritage, and practicality. The majority of musicians speak the language of 12-tet, which provides significant benefits to the community by enabling communication across cultures and genres. Moreover, 12-tet gives access to an incredibly rich library of masterpieces, from church chorales to modern electronic music. One can make a case that music before the early XX century was not written in 12-tet, thus performing Beethoven or Chopin in 12-tet is not historically accurate.[23] However, 12-tet does a good enough job of approximating historic meantone temperaments, so all classical pieces are playable in 12-tet, and the difference in sound is only perceptible to a skilled listener. That naturally brings us to the third point that 12-tet is very handy as a practical tool. Since the 16th century, as philosophical foundations have become increasingly materialistic, practicality has become a persuasive argument, contributing to the popularity of tempered tunings. If there were something more practical than 12-tet, or something desirable that 12-tet could not replicate, it would not get the reach it got.

The practicality of 12-tet is twofold. One benefit is that it suits very well for fretted and keyboard instruments; in fact, the whole process of tempering was introduced into tuning theory to make those instruments sound in tune. The second benefit is that it simplifies the language of music. A modern musician spends countless hours learning to read sheet music, understand how notes work together, and how to express certain feelings and emotions through music. Thus, any simplification on that journey is very welcomed. Meantone tuning made a modern keyboard possible; the layout of white and black keys is a marvel of engineering. It is easy to navigate and comprehend, and it gives so many possibilities to a skilled performer that it has become the go-to tool for every composer. The shift from meantone to 12-tet removed the expressive nuance of different tonalities, but at the same time, it made the language of music more coherent and opened new possibilities for experimentation with modulation.[24]

[21] [8] Chapter 1 "One definition of consonance – the one used in this book – is that two pitches are consonant when their vibrations can be related by small numbers".

[22] [12] Chapter 15: A thumbnail sketch of the history of intonation

[23] In meantone temperaments all tonalities had their own color, as intervals were not uniform as in 12-tet. That is the reason why great composers of the past choose different tonalities for their pieces. Sometimes that choice of tonality leads to more difficult fingering on an instrument, thus demanding a greater skill of the performer. In 12-tet, that color difference between tonalities is lost. So, the only difference that remains is the relative pitch difference and different fingering.

The alternative tuning systems fall short in delivering better practicality than 12-tet. In fact, they add a lot more complexity to the already very complex art of music. Let's consider JI as an example. Instead of learning seven notes, musicians will need to learn different flavors of those notes and start dealing with the arithmetic of simple ratios. Those other flavors must be somehow written down in the notation, so now the complex art of sheet reading becomes even more complicated. Moreover, extra difficulty is added to the instruments used to perform JI music. Keyboards that can approximate intervals of JI are not widely available, and using fretless instruments requires even greater skill and ear training to hit the correct notes consistently. Only the brave souls take that route, and they reap the benefits of JI's unique sound, which is not available to 12-tet musicians.

That said, I believe this critique may no longer be relevant as modern digital tools can address many of those issues. Today, people often listen to recordings instead of attending live shows. Such music is either recorded in a controlled studio environment or entirely digitally produced. That lowers the skill required for performances outside of 12-tet. Most modern musicians are unfamiliar with classical notation because they rely on digital notation, which is quite different. Notes in modern software are represented as a grid of MIDI. It seems likely that certain note flavors can be encoded much more easily on a screen than on paper. Additionally, computers can handle mathematical calculations, relieving musicians of the need to deal with numbers.

We can conceptualize that instruments and notation make an interface to tuning, which in turn is an interface to music. In the case of 12-tet, piano keyboard and classical notation are interfaces to 12-tet, which in turn is an interface to European music. A lot of times, you can substitute one of those elements without the need to change the other two. For example, you can substitute piano for guitar and notation for tabs, while keeping 12-tet and European music intact. One can think of many similar examples; however, as we go up the interface hierarchy, it becomes more difficult to keep something intact. For example, a slight change from 12-tet to meantone temperaments does tolerate keeping keyboard and notation, but it limits the scope of European music, as some tonalities are inaccessible in the meantone tuning that are accessible in 12-tet. And if you change the European music to traditional Middle Eastern music, you must consider switching both the tuning and the instrumentation to achieve a faithful performance.

[24] Modulation is the process of switching from one tonality to another. Different transitions have their own specific character that can be used to express certain feeling. In meantone tunings some tonalities are unplayable, thus one cannot modulate to them. In 12-tet there is no limitation on modulations as all tonalities are equal.

The great thing about 12-tet is that it can approximate a wide variety of music. It can even approximate the music of other cultures, such as Middle Eastern music, but with varying degrees of success and an inherent Europeanization of sound. Sometimes, however, the music is just too foreign for 12-tet. There are attempts to perform Gamelan using 12-tet, but, in my opinion, they do not resemble real Gamelan at all. That property of 12-tet as a king approximator has to be recognized and put in its proper place; it is a tool to achieve a specific sound and serve a specific purpose, but it is not a universal truth of music.

Having that mindset, we can start thinking about tuning as a tool that serves us. In some cases, JI's distinct, stable sound would better suit a composition than 12-tet. The problem that keeps many musicians from experimenting with tunings is a lack of a proper interface (instrument and notation) for tunings other than 12-tet. To this day, the piano keyboard remains a dominant tool for composers, and it inherently leads to 12-tet. That coupling of keyboard and 12-tet was noticed by Harry Partch in the middle of the 20th century, which set him on a quest to build new musical instruments that interface with different tunings and produce different music. The same coupling exists between 12-tet and modern notation that powers all electronic music – the MIDI protocol. By design, MIDI was explicitly developed as a digital interface for 12-tet. So, for the 12-tet to have its fair place as a keyboard-specific interface for European music, and for other tunings to become relevant, a new interface to tuning has to be developed. That interface should enable effortless experimentation with tuning and, ideally, simplify access to European music compared to the 12-tet and piano keyboard, thereby helping it gain popularity. But what tuning theory should such an interface target, if any?

## 2.4 Methodology for a new theory

The question of whether natural, non-subjective tuning exists has baffled theoreticians for thousands of years. It is of primary importance when building an interface for music. If natural tuning does not exist, one is free to choose any tuning to suit their needs. In that case, an interface for tuning should be as flexible and non-discriminating as possible so that it does not influence musicians' choice for a particular tuning. Digital technologies are flexible enough to support such an interface. However, I am a firm believer that natural tuning does exist. That belief stems from my belief in the benevolent God and the intelligent design of Creation. Furthermore, I subscribe to what some call a Pythagorean faith [3], a belief that fundamental structures of the Universe, such as natural tuning, can be found using reason and expressed in the language of mathematics. For those who do not share my beliefs, I think that Sethares' success in describing the music of Gamelan and European music within a single theory points towards the existence of natural tuning. How else is that possible if there is no cross-cultural, non-subjective mechanism that determines tuning? In the end, however, one must take the existence of natural tuning on faith.

If natural tuning exists, the interface to music should be built around it. One can search for such a theory in two ways. The first is empirical: by collecting as much data as possible and then inductively generalizing to a set of ideas that describe all the empirical data. The work [1] is a good example of such an approach, as it starts by generalizing empirical findings into three mechanisms of consonance perception[25]. And then determining the proportion each mechanism plays in our perception of consonance to best fit the empirical data. That approach works, but it has a major problem. It does not suggest a natural tuning system. No matter how much empirical data you gather, there will always be some things that contradict it. You will have to endlessly tweak parameters to keep improving the system. Such an approach can only inform the mind of where to look for a natural system and what to pay attention to, but it will never find it.

The only way to find the natural tuning is by suggesting an idealistic theory. Idealistic theory places the origin of consonance, and thus of tuning, outside the material, sensory realm. It starts by abductively postulating a set of axioms from which the rest of the theory and testable predictions are derived deductively. The benefit of such theories is that their axioms are usually very simple mathematical ideas that are timeless and can be transmitted across millennia. To claim the validity of the theory, one can choose the way of the Pythagoreans, postulate one's reasoning, and claim the correctness of the theory by reasoning alone. Or the way of Ptolemy, to judge correctness by how well predictions of the theory fit the empirical data. I believe that Ptolemaic methodology is the same one used in physics from the time of Newton and is responsible for the staggering achievements in the description of material reality.

[25] Cultural familiarity, spectral interference and harmonicity / periodicity.

As we discussed before, there is no modern theory of consonance that describes all empirical data. In the table below, one can see what is missing in each family of theories.

Table 2.1 – Limits of applicability of current tuning systems

| | Conventional harmony | Consonance of inharmonic sounds | Provides tuning system for pure tones (sine waves) |
|---|---|---|---|
| Just Intonation | ✓ | ✗ | ✓ |
| Harmonicity / periodicity | ✓ | ✗ | ✓ |
| Sethares spectral interference | ✓ | ✓ | ✗ |

The natural tuning theory should check all the boxes and, hopefully, yield testable predictions beyond what is stated here. Such a theory is free of subjective factors, as it does not treat consonance as a sensory phenomenon. Senses detect consonance as an eternal mathematical reality and inform the mind about it. The mind, depending on its state, can have a different disposition relative to that reality by admitting or rejecting it.

In the past, attempts to formulate such a theory were made by using the Idea of proportion as the source of consonance. In the same way our eye recognizes the beauty of correct proportions in architecture or visual arts, our ear recognizes it in sound. The beauty of properly proportioned sounds is what is perceived as consonance. That is the basis for Pythagorean, Ptolemaic, Zarlino's, and JI tuning systems. In this work, I suggest another mathematical approach that can lead to natural tuning theory. I propose that consonance is not codified in the mathematics of proportions, but rather in the mathematics of sets. Using that approach, we can quantify a value of consonance for a given interval that accounts for both spectral interference and harmonicity within a single framework. That leads me to assume that the proposed theory checks all the boxes in Table 2.1, though in this paper, I do not present a study that definitively proves it.

I believe that the proposed approach is promising enough to be implemented in practice. However, I do not suggest that my theory is a natural tuning system in its perfected state. More research and critical analysis are needed. My intuition suggests that there is a more general mathematical idea behind consonance than the one proposed in this work. I share my thoughts and conclusions as a possible stepping stone to inspire others to find what I can't, or to use what I discovered to make some good music.

# 3 Mathematical representation of sound

Conventional tuning theories are built on the assumption that musical sounds have a pitch. Such sounds can be characterized by the frequency of a pitch. Thus, a single number is enough to represent a musical sound. In reality, sound is a much more complex phenomenon, and to capture that complexity, we have to find a more appropriate mathematical representation.

## 3.1 Physics of waves

From the physical standpoint, sound is a longitudinal pressure wave that propagates in a medium, usually air. To create such a wave, a body has to hit a medium to create a zone of high pressure that will propagate with constant speed $v_s$ (speed of sound). A single hit creates a single shock wave that will be perceived as a short click. To make a continuous sound like a voice, the body should vibrate, aka repeatedly hit a medium for the duration of the sound (Figure 3.1.a). The rate at which the body hits the medium determines the pitch of a sound.

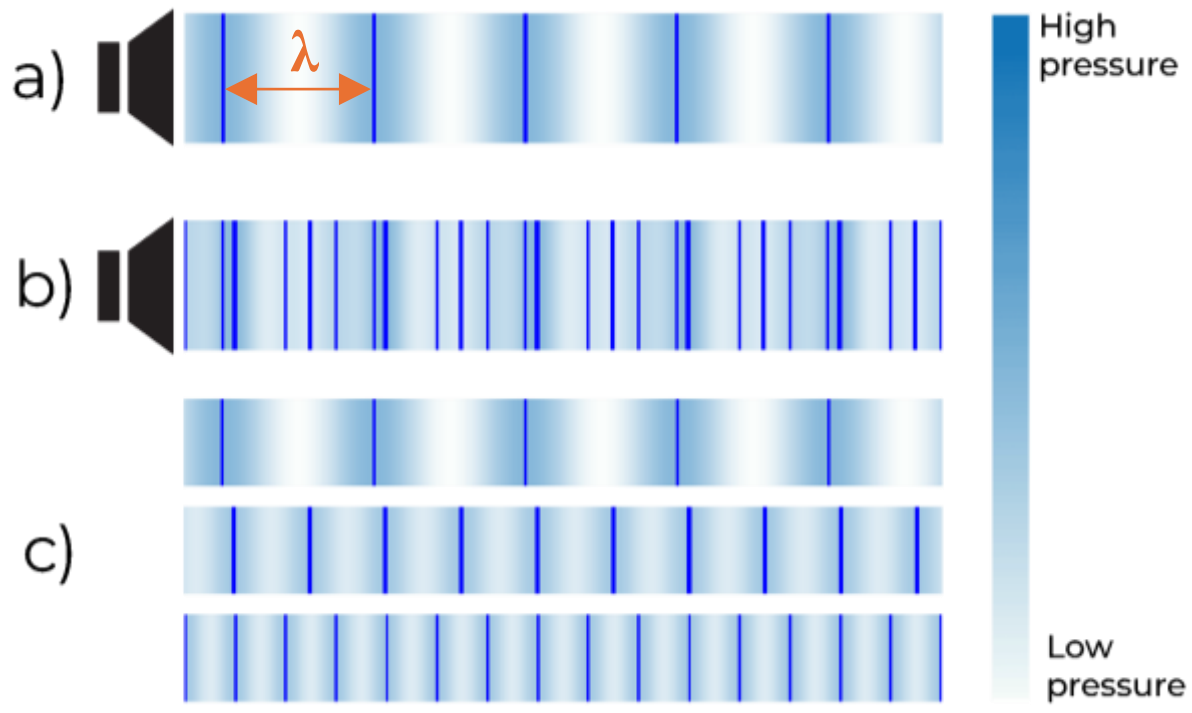

Figure 3.1 – Different types of pressure waves that can propagate in a medium. a) pressure wave created by a speaker that continuously hits a medium with a constant period $\lambda$. Blue lines mark the spot where the hit occurred. b) a complex pattern with which the speaker hits a medium. c) a set of simple periodic hit patterns that make up the complex pattern in b). Notice that all the blue lines in b) have a corresponding blue line in c).

The simplest pattern (oscillation) with a constant pitch is when hits occur at a constant rate. More complex patterns are certainly possible, like the one in Figure 3.1.b. Interestingly, one can notice that three elementary oscillations make up that complex pattern (Figure 3.1.c). In mathematics, representing a signal as a sum of elementary oscillations is called the Fourier transform. Different waves will have different sets of oscillations, but any wave can be described this way. Thus, one might say that the sound in Figure 3.1.b has several pitches sounding at once.

From now on, we will refer to such elementary periodic oscillations, as shown in Figure 3.1.c, as **partials** of a sound. Each partial can be characterized by a number $\lambda$ – wavelength of oscillation. Wavelength is the distance in meters in which the waveshape repeats (Figure 3.1.a). Knowing the speed of sound in a medium $v_s$ and the wavelength $\lambda$ , one can find the frequency of oscillation $f$ using the following formula:

$$f = \frac{v_s}{\lambda} \quad 3.1$$

Frequency represents the speed of oscillation, or how many hits are happening per second. The inverse relationship between $\lambda$ and $f$ means that low frequency instruments, like bass guitar, produce sounds with partials of longer wavelength than high frequency instruments like flute. Because each medium has a unique speed of sound, the same body vibrating with the same frequency will generate waves of different wavelengths in different mediums. Thus, frequency is a better characteristic of a vibration as it is medium-independent.

The remarkable thing about human biology is that a part of the inner ear called the cochlea performs a Fourier transform of the sounds that enter it before sending information to the brain [4]. So rather than hearing a complex sound like in Figure 3.1.b, the brain works with a set of partials like in Figure 3.1.c. Moreover, most musical instruments (including our voices) create complex vibrations containing dozens of partials even when a single pitch is being played. This gives us a clue that using a single number to represent a musical pitch may lack important information.

## 3.2 Sound as a set of frequencies

One of the most fundamental concepts in mathematics is the concept of a set. Informally, one can define a set as an unordered collection of unique mathematical objects, such as numbers, vectors, other sets, etc. Sets are denoted by capital letters, and their elements are specified in curly brackets, like so: $S = \{ 2, 7, 6, 1.33 \}$. Here $S$ is a set of numbers: 2, 7, 6, 1.33. Another way to write a set is by stating properties that define its elements: $S = \{ s \mid s \in \mathbb{N},\ s < 5 \}$. This reads like this "The set $S$ is a set of all $s$ such that $s$ is a natural number and is less than 5.", which means that $\mathrm{S} = \{1, 2, 3, 4\}$. Refer to Table 8.1 and Table 8.2 in the supplementary materials for the list of symbols and operations on sets used in this article. Operations in Table 8.2 are specific to this paper and are not used in the common set theory.

As we previously discussed, real musical instruments produce complex vibrations that consist of many partials, each with a unique frequency. Thus, rather than using a single number to represent the pitch of a sound, we can describe it as a set of frequencies of all its partials:

$$F = \{f_1, f_2, \dots, f_k\} = \{f_i \mid i = 1, 2, \dots, k\} \quad 3.2$$

Here, $F$ is a set that represents a sound that is made of $k$ frequencies $f_i$. The example of what that means is shown in Figure 3.2. The complex wave in Figure 3.2.a shows how the pressure on the Y-axis varies with time on the X-axis, which is called the **time-domain** representation. It can be transformed into the **frequency-domain** representation using the Fourier transform (black line in Figure 3.2.b). The amplitudes and the frequencies of individual partials that make up the sound are marked by the peaks on the Fourier transform graph. However, there is always implicit uncertainty in the Fourier transform derivation of partial frequencies and their amplitudes. The precision with which those amplitudes and frequencies are derived depends on the input data. The exact frequencies that build up the sound in Figure 3.2.a were known in advance and are shown as colored vertical lines in Figure 3.2.b. We will present the frequencies and amplitudes of partials in the frequency domain using vertical lines from now on. The time domain representation of partials in Figure 3.2.b is shown in Figure 3.2.c. Summing all the graphs in Figure 3.2.c yields exactly the graph in Figure 3.2.a.

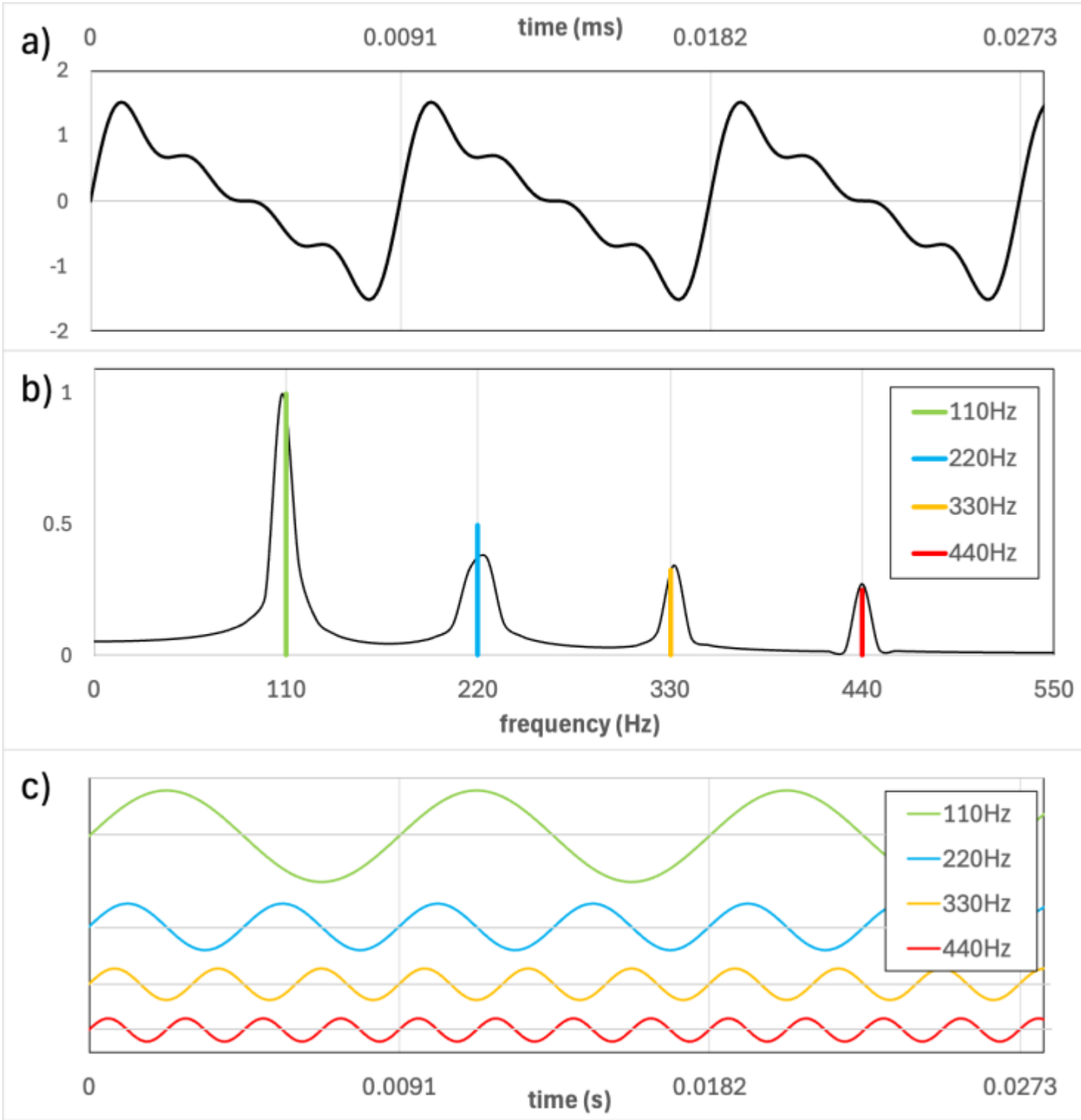

Figure 3.2 – Demonstration of how a complex wave can be seen as a collection of elementary oscillations (partials) in both time and frequency domains. a) The time-domain representation of a complex wave. b) The black line is the Fourier transform or a frequency-domain representation of the wave in a). Colored vertical lines represent frequencies and amplitudes of partials that comprise the wave in a). c) The time-domain representation of individual partials that comprise the wave in a), colors correspond to b).

As we saw, every partial in a sound is characterized by 2 numbers, frequency and amplitude. However, for simplicity, we will not include amplitude into consideration and limit our analysis to frequencies only. Thus, the set of frequencies that corresponds to the sound in Figure 3.2 is following:

$$F = \{110, 220, 330, 440\} \quad 3.3$$

## 3.3 Harmonic sets

There are as many different sets of frequencies as there are sounds. Some sets may appear structured; others may seem random. The majority of musical instruments produce **harmonic** sounds in which each frequency is an integer multiple of a **fundamental** frequency. The sound in Figure 3.2 is an example of such a sound because every frequency in Formula 3.3 is an integer multiple of $110\ Hz$, which is a fundamental frequency of that sound. From now on, we will denote fundamental frequency as $f_F$. Because harmonic sounds are very common in music and follow a particular structure, we can use the definitions in Table 8.2 to rewrite the set $F$ more compactly $F = 110 \cdot \{1, 2, 3, 4\} = 110 \cdot \mathbb{N}_4$. In general, we can define harmonic sounds as those that can be described by the following formula:

$$F = f_F \mathbb{N}_k \quad 3.4$$

One of the features of harmonic sounds is that they are perceived as a single pitch with the frequency of $f_F$. The number of frequencies $k$ determines the timbre of the sound and depends on many factors, such as the intensity with which the note is struck. The larger the number $k$, the richer and brighter the timbre is. In reality, the number $k$ changes over time as high frequencies tend to decay faster than low frequencies. We will not pay attention to that dynamism and suppose that a typical note played on a musical instrument has a fixed fundamental frequency and a fixed small number of partials $k$.

It is clear from Formula 3.4 that fundamental frequency is an elementary vibration from which all other frequencies in a set are made. In mathematics, the number that serves that purpose is known as the **greatest common divisor** (gcd)**,** thus it can be written that:

$$f_F = \gcd(F) \quad 3.5$$

The geometric meaning of gcd is shown in Figure 3.3. One can see that gcd is sort of a building block for a set of numbers.

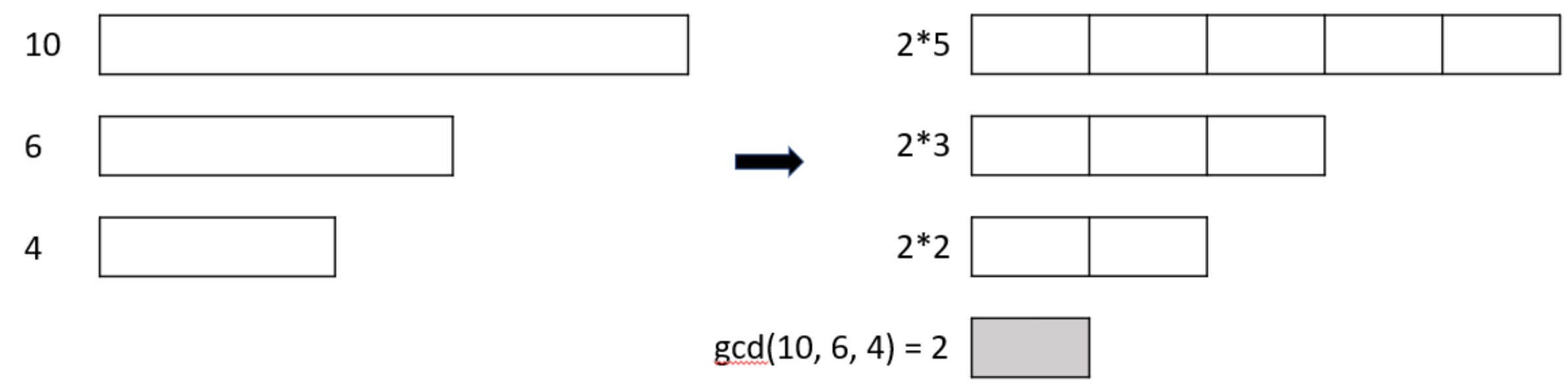

Figure 3.3 – Geometric meaning of the greatest common divisor (gcd). On the left, there are 3 numbers: 10, 6, and 4, each representing a unit of length. On the right same numbers are the results of stacking gcd(10, 6, 4) = 2 a whole number of times. It is evident that gcd is a number that can be viewed as a unit length from which other numbers can be made. Greatest common divisor does not have to be a whole number.

The key feature of the harmonic sounds is periodicity. An example of that is shown in Figure 3.2.a, where the waveform repeats with a period of $T_{total} = 9.1\ ms$. The reason for that periodicity is purely geometrical. At some point, the periods of all constituting partials coincide, which resets the wave to its initial state and causes the pattern to repeat. That mechanism is evident from comparing the oscillation periods in Figure 3.2.a and Figure 3.2.c.

In mathematics, the smallest number to which some other numbers fit a whole number of times is called the **least common multiple** (lcm). It can be used to find the period of the total wave from the set of periods of its partials. To do it, we need to convert a set of frequencies to a set of periods:

$$T(F) = \left\{\frac{1}{f_i} \mid f_i \in F\right\} \tag{3.6}$$

and find the lcm

$$T_{total} = \mathrm{lcm}(T(F)) \tag{3.7}$$

Let's notice that the period of a total wave in Figure 3.2.a is the same as the period of the fundamental in Figure 3.2.c. That is a consequence of the reciprocal nature of frequency and period, and of the geometric meaning of gcd. Let's look at the example of a harmonic set $F = \{f_F, 2f_F, 3f_F, \dots, kf_F\}$. According to Formula 3.6 the corresponding set of periods will be $T(F) = \left\{\frac{1}{f_F}, \frac{1}{2f_F}, \frac{1}{3f_F}, \dots, \frac{1}{kf_F}\right\}$. We can substitute the period of fundamental $T_F = \frac{1}{f_F}$ and write $T(F) = \left\{T_F, \frac{T_F}{2}, \frac{T_F}{3}, \dots, \frac{T_F}{k}\right\}$. Because each consecutive period in the set is a quotient of $T_F$ on a natural number, we can write $T_F = \text{lcm}(T(F))$. On the other hand, $T_F = \frac{1}{f_F}$. Thus, the total period of the wave can be found using gcd instead of lcm:

$$T_{total} = \frac{1}{\text{gcd}\,(F)} \tag{3.8}$$

This way, the role of fundamental is twofold. On one hand, the frequency of a fundamental is the building block for the rest of the frequencies in a set. On the other hand, it has the same period as the total wave. That means that if different sets have the same fundamental, they will have the same period and vice versa.

## 3.4 Phantom fundamental, inharmonicity, and periodicity

Not every sound has a real fundamental. Let's take a set $F = \{440,\ 550\}$. Using the Formula 3.5 we get $f_F = \text{gcd}(\{\,440, 550\,\}) = 110$. But 110 Hz is not an element of the set $F$, or in mathematical notation $f_F \notin F$. In such a case, it is said that the sound has the **phantom fundamental** of 110 Hz. The phantom fundamental has the same properties as **a** real fundamental. It is also the building block of the rest of the frequencies in the set, and it has the same period as the total wave (Figure 3.4). The surprising thing is that the perception of the pitch of the total wave does not change if the sound has a real or a phantom fundamental. Meaning that sounds in Figure 3.4 are perceived to have the same pitch even though the phantom fundamental is not directly present in the sound.

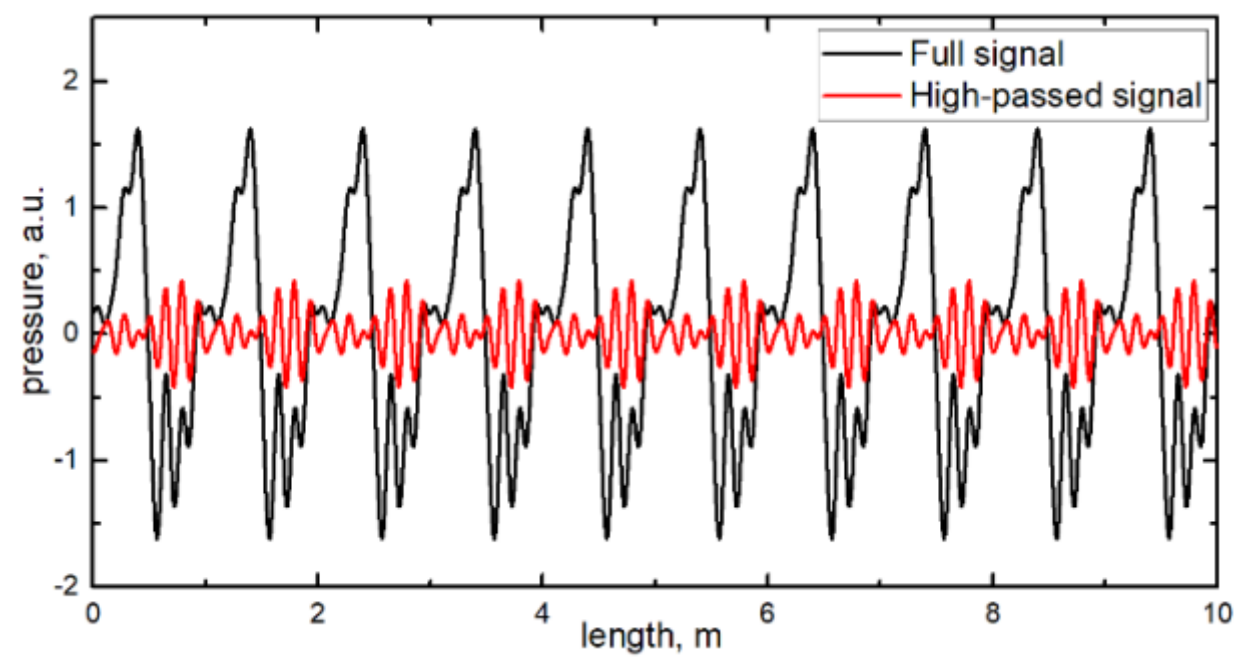

Figure 3.4 – The black line is the waveform of the harmonic spectrum with seven partials, including the fundamental frequency. The red line shows a waveform of the same spectrum but with the first four partials removed. Note that the periods of the two waves are the same. The period of the black line is the period of its real fundamental, and the period of the red line is the period of its phantom fundamental, which are equal in this example.

Strictly speaking, for some sets the gcd does not exist. If a ratio between any two frequencies in a set is irrational, no matter how small the building block (gcd) we pick, we won't be able to build up both of those frequencies with it. In mathematics, the numbers $a$ and $b$ are called **commensurable** if their ratio $a/b$ is rational. Geometrically, that means that for such numbers there is a common unit of length $c$ that can be used to measure both $a$ and $b$, $a = nc$ and $b =$ $mc$ where $n, m \in \mathbb{Z}$. That is a necessary and sufficient condition for the existence of $\gcd(a, b)$. If $a/b$ is irrational, numbers $a$ and $b$ are called **incommensurable** and $\gcd(a, b)$ does not exist**.** Thus, we can say that if the set $F$ is commensurable, meaning all frequencies in $F$ are commensurable, then $\gcd(F)$ exists.

If a set $F$ is such that there is no $\gcd(F)$, then the corresponding sound does not have a fundamental, and the wave of such sound is not periodic. While such a case is possible in mathematics, it is not possible in practice because achieving an irrational ratio between two frequencies requires infinite precision. Nevertheless, some combinations of partials can produce phantom fundamentals with wavelengths many thousands of times larger than those of the constituting partials. That pushes the fundamental outside the hearing range and makes the sound appear non-periodic.

Figure 3.5 demonstrates how the period can change with the structure of a set. The sounds in Figure 3.5.a and Figure 3.5.b have simpler ratios between partials, resulting in smooth, stable sounds and high periodicity. The seemingly similar set in Figure 3.5.c has a much longer period and thus will be perceived as an unstable sound with noticeable volume variations.

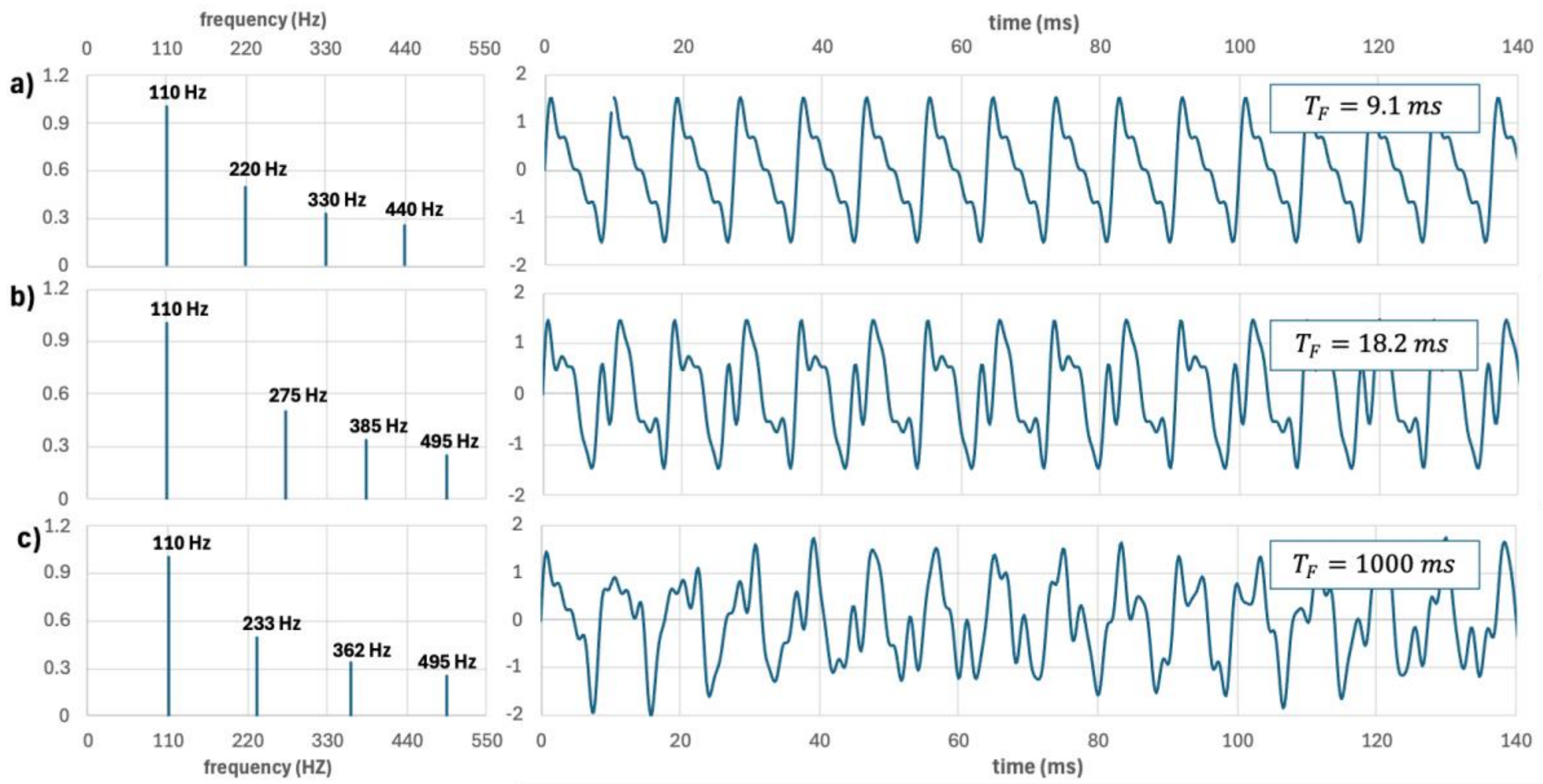

Figure 3.5 – The relationship between a set of frequencies and the period of a sound wave. a) Harmonic spectrum with fundamental $f_F = 110\ Hz$ with a period of $T_F = 9.1\ ms$. b) Inharmonic sound with second, third, and fourth partials shifted to higher frequencies. That shift results in the emergence of a phantom fundamental $f_F = 55\ Hz$ with the period $T_F = 18.2\ ms$. c) Inharmonic sound that may seem more uniform in the frequency domain than the sound in b), but with the significantly lower phantom fundamental that is outside of a hearing range $f_F = 1\ Hz$ with a period of $T_F = 1\ s$. Such a long period results in a perception of instability in the overall sound.

The sounds that deviate from a harmonic set, such as in Figure 3.5.b and Figure 3.5.c, are called **inharmonic** sounds. Britannica defines inharmonicity as "deviation of the frequencies of the harmonics from the exact multiples of the fundamental". That assumes a different definition of fundamental than the gcd of a set of frequencies, as in the case of gcd, frequencies of the harmonics cannot deviate from integer multiples by definition. What is called a fundamental in modern empirical tradition is the most pronounced low-frequency partial of a sound that is perceived as a pitch of that sound. From that perspective, because all sounds in Figure 3.5 are perceived as having the same pitch, they have the same fundamental frequency of 110 Hz. The difference in spectrum is perceived as a difference in timbre.

Which definition of fundamental to pick for a theory depends on the philosophical tradition that you submit to. I believe that the strict mathematical definition of fundamental reveals a geometrical meaning behind the relationship of frequencies within a sound. Some frequencies fit together forming periodic patterns, while others don't. That will help us mathematically define consonance in Chapter 4.3 and use that knowledge to build a tuning theory. The fact that perceptions may deviate from pure mathematics should be accounted for separately by modelling perceptions within the idealistic framework. Building such a model is not the task of the current paper.

## 3.5 Defining notes as sets

As mentioned in Chapter 2.3, the new tuning framework must be as convenient as possible to have a chance of success. Thus, it is important to have a system of pitches that is familiar yet more flexible than the common one. Modern 12-tet tuning was developed for keyboard instruments, and, like the piano keyboard, it has 12 notes that repeat every octave. Each note has a name (A4, C#5, etc.) and its corresponding frequency, forming a very strict grid.[26] In reality, acoustical musical instruments, including pianos, cannot be tuned perfectly to that grid, so some pitch deviations are inevitable. If a deviation of note from that grid is significant enough to be picked up by a human ear, the note is considered to be out of tune. For example, the frequency for C#5 on a piano should be 554.37Hz. If we tune that note to 550Hz instead, which is a perceivable difference, the instrument will be considered out of tune.

[26] The grid of 12-tet pitches has the note A4 tuned to exactly 440Hz and all other notes are determined by shifting that frequency the correct number of semitones away. For example, C#5 is 4 semitones higher than A4, thus it will have a frequency $440 \cdot 2^{\frac{4}{12}} \approx 554.37\ Hz$.

Such tuning of instruments to a predefined grid of pitches is how most modern musicians tune their instruments. Moreover, modern tuning practice does not assume that there are other notes outside of that grid. If you follow it, you are in tune; if you don't, you are out of tune. That is a reality we have to face, and to make the system more accessible to modern musicians, I suggest sticking to established 12-note notation while easing the strictness of note-frequency correspondence. We will assume that a note corresponds to a range of frequencies, rather than a specific frequency. A range for a note is calculated as $\pm 50$ cent difference from pitches of the 12-tet grid. This way, if a pitch is within 50 cents of 554.37Hz, it will be considered C#5. The complete list of frequency ranges is presented in Table 8.3.

Having stated that, we can devise a way to mark which note a given set corresponds to. Let's consider the sound in Figure 3.2 as an example. The fundamental frequency of that sound is $110\ Hz$, which according to Table 8.3. corresponds to the note A2. So instead of using generic $F$ to refer to the set, we can use the note's name instead. We can also use a subscript index to specify the number of partials $k$ that the note has. This way, the set $F$ in the Formula 3.3 can be rewritten as $A2_4$, which clearly indicates that $F$ is the A2 note and is made of 4 harmonic partials.

Because in the 12-tet system, the intervals between notes are fixed, defining the frequency of one note defines the frequencies of all notes. In our system, the frequencies of notes are defined by a relative interval from a note with a known frequency. This way, if $A2_4 = 110 \cdot \mathbb{N}_4$, we can define $E3_4 = \frac{3}{2} A2_4$, which is just fifth up from $A2_4$, or a $C\#3_4 = \frac{5}{4} A2_4$, that is just major third up from it.[27] That is a very convenient way of notation within the set-theoretic approach that will be used in the following chapters. It also allows precise notation of sets corresponding to chords. For example, the pitches we defined previously form a just major triad. The set for that triad is the union of all the notes that form it $A2\text{-}dur_4 = A2_4 \cup C\#3_4 \cup E3_4$.

Set theory seems like a suitable language for describing a tuning theory. The most important takeaway from this chapter is that a pitch is better represented as a set of numbers rather than a single number. That realisation is a key to the mathematical definition of consonance that will be proposed in the next chapter.

---

[27] Notice, that once we defined $E3_4$ we can define $C\#3_4$ relative to it, instead of $A2_4$. $C\#3_4$ is a just minor third down from $E3_4$ or $C\#3_4 = \frac{5}{6} E3_4$, which is exactly the same as $C\#3_4 = \frac{5}{4} A2_4$. There is no correct way of doing such definitions, it depends on the context of what you are trying to achieve.

# 4 Measures of consonance

In this chapter, I will suggest two measures of consonance for arbitrary sets of frequencies that represent sounds. These measures are harmonicity and affinity. Affinity measures how closely related different sounds are and, consequently, how well they fit together when sounded simultaneously. Harmonicity measures the extent to which a set of frequencies forms a single pitch. Both measures are numbers between 0 and 1, where 1 means perfect consonance and 0 absolute lack of consonance. I will also compare those two measures with similar theories from the literature.

## 4.1 The definition of consonance

To build mathematical measures of consonance, let's recall the definition of consonance given in Chapter 2.1. We can restate it as follows.

Consonance is an observation that some sounds, when played together, sound:

1. smooth and stable,
2. harmonious as if they fit together forming one larger whole.

Here, we divide the definition into two parts. The first part describes the sensory aspect of consonance, an observation of smoothness and stability. The second part relates to the philosophical problem of unity and multiplicity, the mysterious phenomenon when many become one.

Within an idealistic framework, the second aspect of consonance must be held as primary, from which the first, sensory, aspect should unfold. Thus, let's state the definition of consonance for which we will seek a mathematical description. I suggest it should sound like this:

"Consonance - is a phenomenon when multiple sounds fit together forming one greater harmonious whole."

A mathematical expression of this idea can serve as a measure of consonance. Successful measures, when applied in practice, should produce sounds that are perceived as smooth and stable. Many such measures can be proposed, but the simplest ones that best match observations are preferred.

## 4.2 The affinity of sets

In Chapter 3, we concluded that the appropriate mathematical object to describe a sound is a set of frequencies. One of the aspects of our definition of consonance states that consonant sets are the ones that fit together. We can say that sets fit together if they share common elements. Mathematically, that commonality can be measured as the relative size of a shared fraction between different sets. If the proportion of shared elements is big, then sounds are closely related, and if it is small, then not so much. We will call that measure of shared elements **affinity**[28], and it will serve as a first measure of consonance. The simple way to quantify affinity is the following formula:

$$\alpha(F, F') = \frac{|F \cap F'|}{\min(|F|, |F'|)} \qquad 4.1$$

Here $\alpha(F, F')$ is the affinity of sets $F$ and $F'$. Both of those sets are entirely arbitrary and thus can represent any sound, harmonic or inharmonic, a single note or a complex chord. In the numerator, we have the size of the intersection of $F$ and $F'$, or the number of shared frequencies. If there are no common frequencies between $F$ and $F'$, then the intersection is an empty set $F \cap F' = \emptyset$. The size of the empty set is zero $|F \cap F'| = 0$, which leads to a zero value for the affinity $\alpha(F, F') = 0$. That case signifies an absolute lack of consonance between $F$ and $F'$. Graphically, that case is represented in Figure 4.1.a.

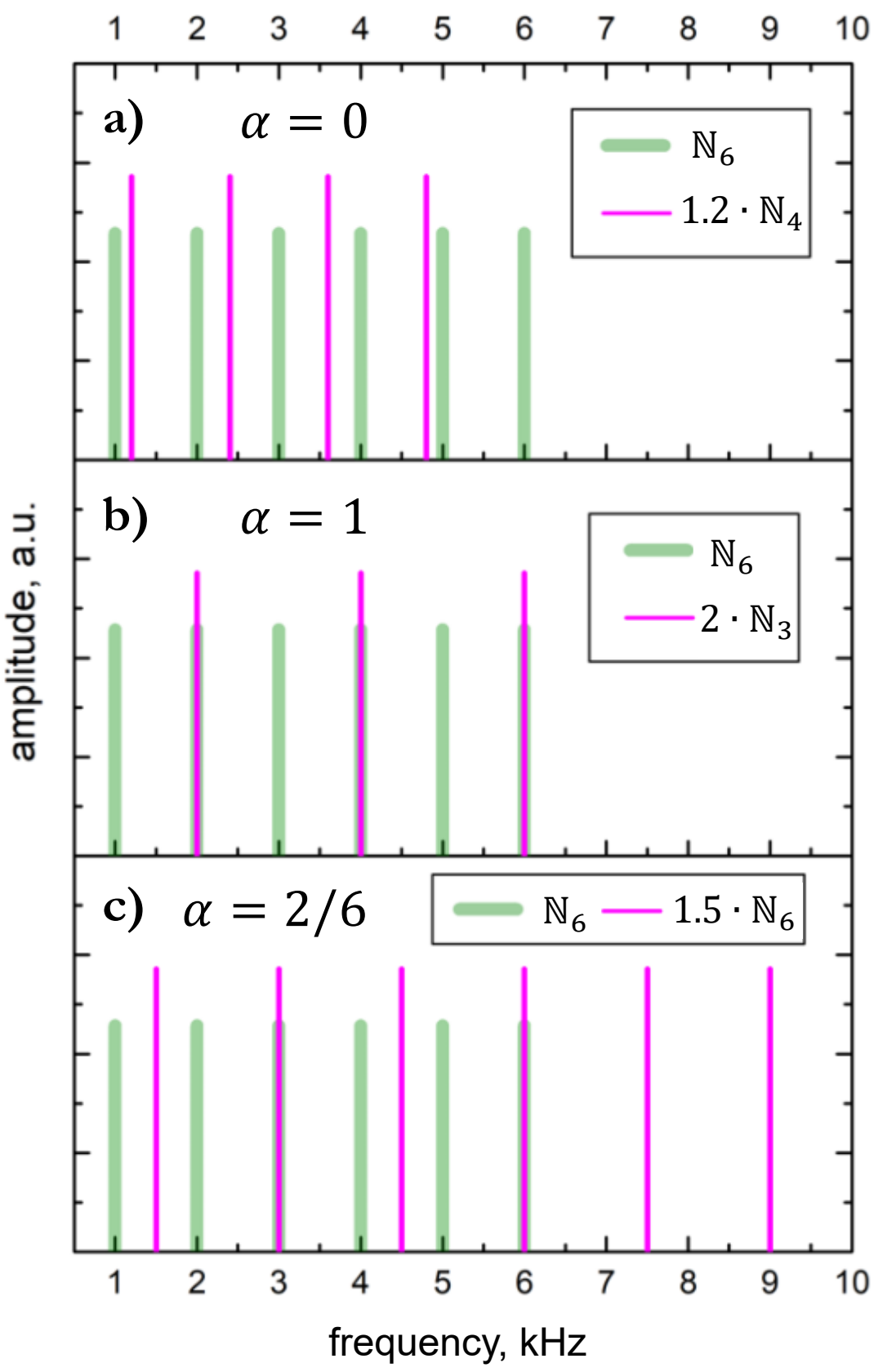

Figure 4.1 – Demonstration of how different sets can have different values of affinity. a) The sets $\mathbb{N}_6$ and $1.2 \cdot \mathbb{N}_4$ do not share any frequencies, thus having a 0 value of the affinity. b) A set $2 \cdot \mathbb{N}_3$ is a subset of $\mathbb{N}_6$, thus the affinity is equal to 1. c) The realistic case of perfect fifth, two sets $\mathbb{N}_6$ and $1.5 \cdot \mathbb{N}_6$ have 2 partials in common, which makes the affinity to be $2/6$.

If there are common elements between sets, then $|F \cap F'| > 0$ and the affinity is not 0. To determine the exact value, we have to calculate the denominator, which is the size of the smallest of the two sets $F$ and $F'$. It makes sense to relate the size of the intersection to the smallest set, since this way the maximum value of affinity we can get is $1$. That case is illustrated in Figure 4.1.b, where the smaller set is a subset of the larger one. Let's see how we get the value for affinity in that case. If $F' \subseteq F$, then in the numerator, the size of the intersection will be the size of the subset $|F \cap F'| = |F'|$. As for the denominator, the size of a subset is always less or equal to its superset, thus $\min(|F|, |F'|) = |F'|$. Consequently, the value of affinity becomes $\alpha(F, F') = \frac{|F'|}{|F'|} = 1$, which means the perfect consonance between two sets.

[28] Affinity as a measure of consonance, should not be confused with the notion of the affinity in music theory that refers to scales and scale degrees in tonal music. Those are different notions that may or may not have a deeper connection between them.

In practice, most musical intervals have an intermediary value between 0 and 1, as shown in Figure 4.1.c. That value depends on the number of elements in the smallest set and the size of the intersection. Affinity is a very simple measure of consonance that gives us a numeric value by which we can check the consonance of different combinations of sounds. In the next chapter, we will see how it can be used to build a tuning system.

The idea behind affinity is related to the theories of spectral interference [4], [2], though it has some crucial differences. Sethares' work is based on the limited resolution of the human auditory system, empirically determined in [5]. Using results of that work, Sethares proposes a way to calculate a dissonance value for an arbitrary set of frequencies. If the ear can resolve all frequencies in a set, then the dissonance is low. If many frequencies cannot be resolved, the dissonance is high. Sethares applies that idea to tuning theory by calculating dissonance curves and identifying intervals with relatively low dissonance. The dissonance curve is calculated for two sounds, one of which is a static sound $A$ (a note or a chord), and the other is a sound $B$ (can be the same as $A$) that we want to add to $A$, and that can be transposed on the interval $t$. This way, we can calculate the dissonance value of the combination of $A$ and $tB$ for each value of $t$. The result of that procedure is a curve with maxima and minima, called a dissonance curve, an example of which is shown in Figure 4.2. Sethares proposes that intervals that mark local minima of dissonance are reasonable to include in tuning. He notices that there are two types of minima: sharp minima at which partials of $A$ and $tB$ coincide, and broad minima where partials of $A$ and $tB$ are easily resolved. From his perspective, both are suitable for inclusion in the tuning.

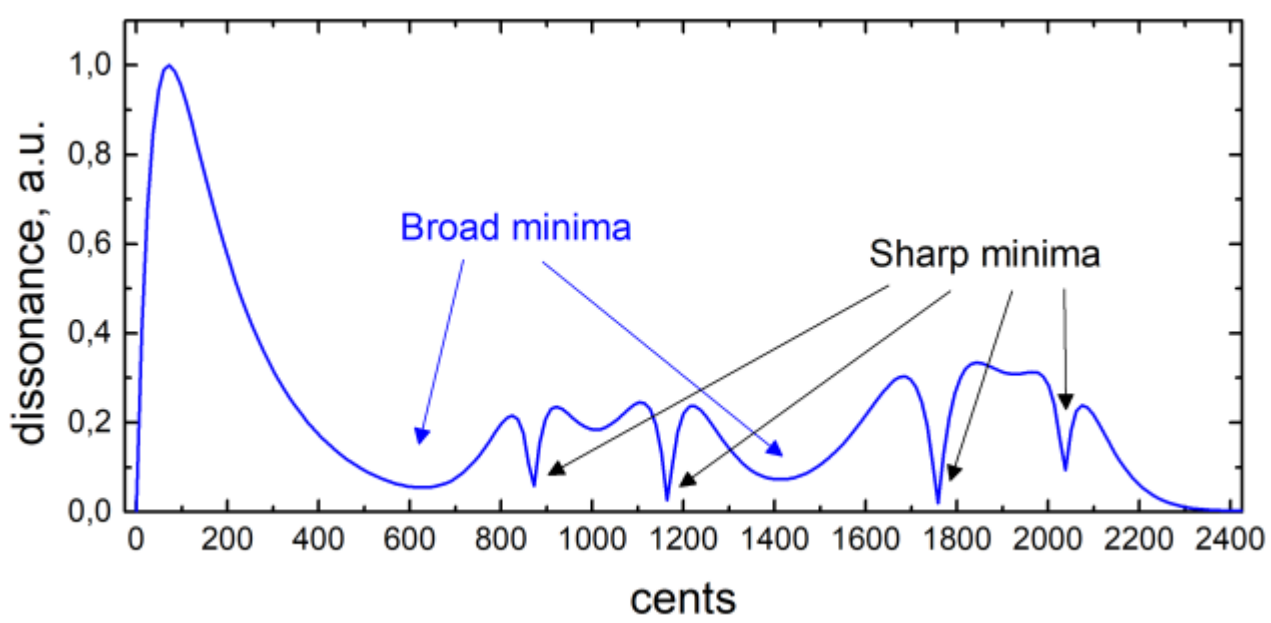

Figure 4.2 – Dissonance curve for two inharmonic sounds that demonstrates two types of dissonance minima: sharp minima, where partials of two sounds coincide, and a broad minima, where dissonance is low due to a large distance between partials, such that the ear is able to resolve them as separate.

The inability of the ear to resolve partials is characterized by a rough sound resulting from a high degree of beating. Thus, the intervals that corresponded to the minima of dissonance have a relatively smooth sound. Sharp minima on a dissonance curve are caused by the coincidence of partials of two sounds, which is precisely the condition for non-zero affinity. As we will see in Chapter 5, tuning derived from dissonance curves can be identical to that derived by using affinity. Thus, we should expect sounds with high affinity to also be relatively smooth-sounding, which satisfies our expectation for the correct consonance measure.

Although spectral interference and affinity approaches share similarities, they differ significantly in their philosophical foundations. Sethares' theory is empirical, whereas the affinity-based theory is idealistic. This results in a couple of advantages for affinity. The first is a dramatic reduction in the computation required to derive tuning. But most importantly, the idealistic theory is independent of biology.

In his work, Sethares argued that capturing broad minima of dissonance is an important feature of dissonance curves. However, I believe that those broad minima introduce a subjective element to his theory. Let's imagine that due to some sickness, the frequency resolution of the inner ear decreases. This means that parameters that describe that resolution in the dissonance curve derivation should also change. As seen in Figure 4.3, the decrease in resolution can lead to the disappearance of broad minima (orange curve) or even to the flattening of sharp minima (red curve). That raises the question: if one becomes deaf, does musical tuning disappear? In idealistic theory, no matter what frequencies are sounding or who is listening, tuning is the same and is determined by the relationship between mathematical objects.

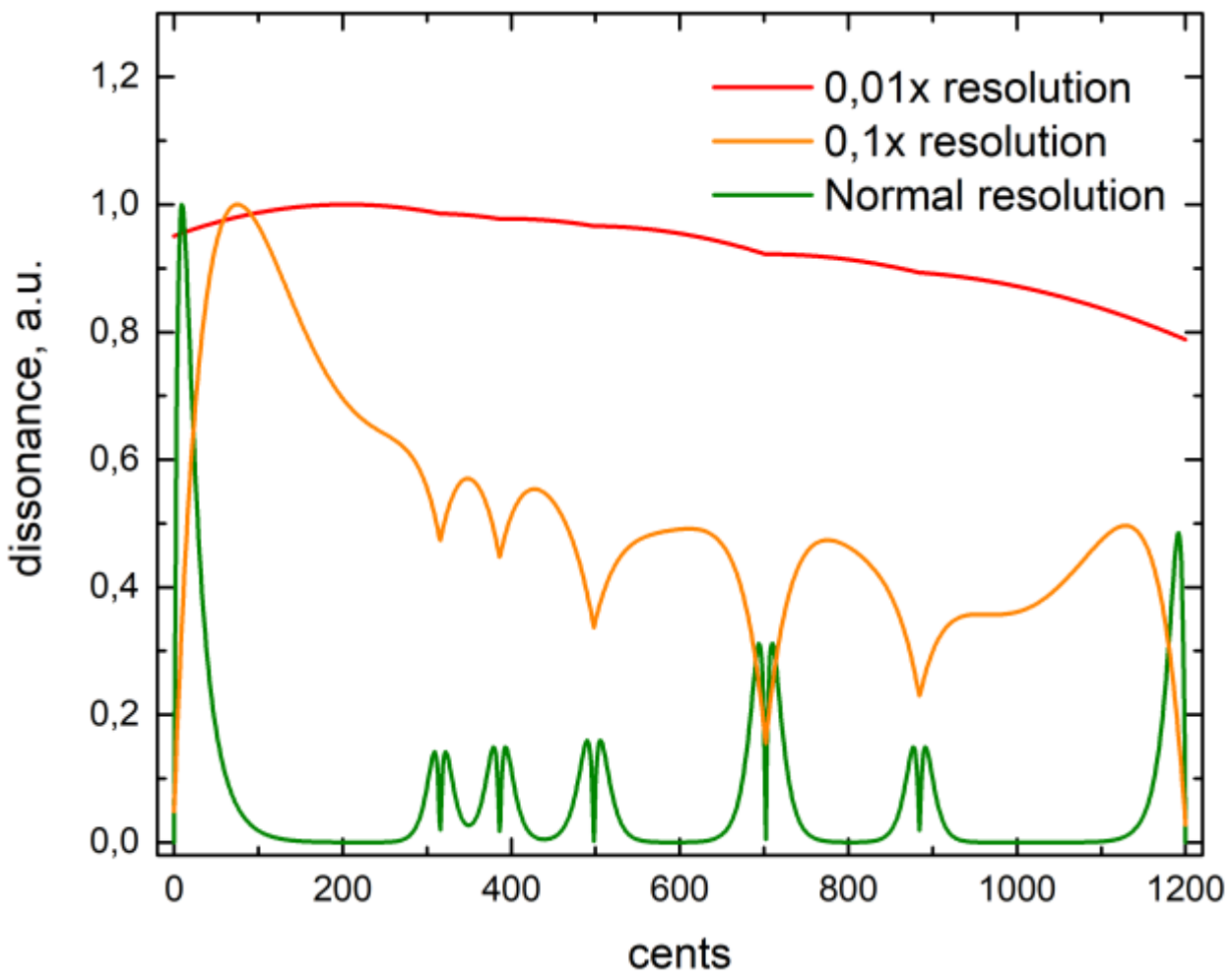

Figure 4.3 – Dissonance curves with parameter $x^*$ varied to demonstrate the lowering of resolution of the auditory system. The green curve represents normal frequency resolution with point of maximum dissonance $x^* = 0.24$ as used in [4]. Lower-resolution curves are obtained by setting that value to 0.03 for the orange curve and 0.003 for the red curve. As the resolution drops, broad minima disappear first, and then sharp minima become much less pronounced.

## 4.3 The harmonicity of sets

The affinity contribution alone does not satisfy the definition of consonance given at the beginning of this chapter. It accounts for sets fitting together, but not for the formation of a greater harmonious whole. To account for that, let's notice that each note is perceived by us as a single pitch even though it is made of many pitches. As we discussed in Chapter 3.3, there is a special arrangement of pitches that we perceive as a single pitch. That arrangement is a harmonic set that is represented by Formula 3.4.

As we discussed before, such sets have many interesting properties and can serve as a powerful abstraction for representing notes played on real instruments. The pattern of integer multiples of the fundamental frequency is commonly known as the harmonic series and is used as a basis for many tuning theories, most notably JI. It is our assumption, backed by common experience, that a set like that in Formula 3.4 represents a single pitch. What is interesting is that any arbitrary set can be compared to how well it resembles a harmonic set. We will call the measure of such resemblance **harmonicity**.[29] The larger the resemblance of a set of pitches to the harmonic set, the greater is the oneness of total sound and therefore, the better they form one greater harmonious whole.

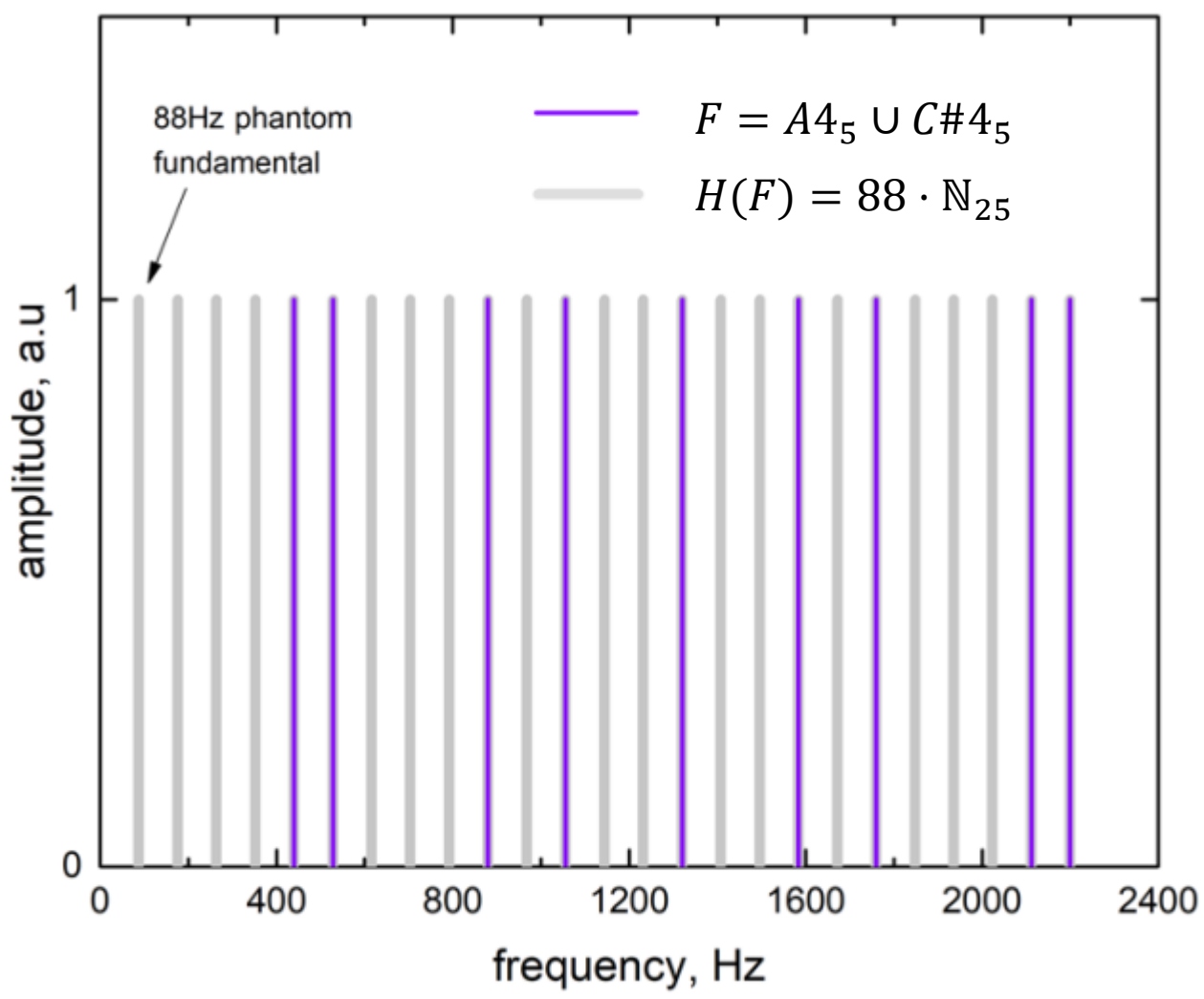

Figure 4.4 – Demonstration of harmonicity as a measure of consonance for the interval of a major third. The violet lines represent a spectrum of a major third interval that is made as a union of the note $A4_5 = 440 \cdot \mathbb{N}_5$ and $C\#5_4 = 5/4 \cdot A4_5$ with frequencies $F = A4_5 \cup C\#5_4$. The grey lines represent the corresponding harmonic superset $H(F) = F2_{25} = 88 \cdot \mathbb{N}_{25}$. One can see that $F$ is the subset of $H(F)$ with the harmonicity value of $\chi(F) = 9/25$. That means that the frequencies that comprise a major third interval point towards a single pitch at 88Hz, but do not form it; they are only part of a larger whole.

To find the harmonicity of a set, we need a harmonic set to which we can compare a given set. Let's remember that for any commensurable set of frequencies $F$ we can find a fundamental frequency as $f_F = \gcd(F)$. As we discussed in Chapter 3.3, all frequencies in $F$ are integer multiples of $f_F$. Thus, we can say that there exists a set $N \subseteq \mathbb{N}$ such that $F = f_F N$. If $F$ is finite, we can find a number $n = \frac{\max(F)}{\gcd(F)}$ that is large enough so that $N \subseteq \mathbb{N}_n$. Knowing that and using the Formula 3.4 we can find a superset of $F$ like so:

$$H(F) = f_F \mathbb{N}_n, \quad n = \frac{\max(F)}{\gcd(F)} \tag{4.2}$$

[29] The similar metrics in literature have different names such as tonalness [4] or harmonic similarity [7]. It can be categorized as pattern-matching harmonicity/periodicity theory according to [1].

The set $H(F)$ is a harmonic set, which means it is a single pitch and $F$ is a subset of $H(F)$. Thus, any set $F$ with commensurable frequencies is a subset (a part) of a single pitch $H(F)$. We will call $H(F)$ a **harmonic superset** of $F$. The meaning behind harmonic superset is illustrated in Figure 4.4, where a set of the just major third made of notes $A4_5 = 440 \cdot \mathbb{N}_5$ and $C\#4_5 = 5/4 \cdot A4_5$, is a subset of a single lower pitch $F2_{25} = 88 \cdot \mathbb{N}_{25}$ with 25 partials. We can use the relative size of a set $F$ to its harmonic superset $H(F)$ as a measure of harmonicity:

$$\chi(F) = \frac{|F|}{|H(F)|} \tag{4.3}$$

Strictly speaking, not every set has a corresponding harmonic superset. If $F$ is incommensurable, then there is no gcd (F), and thus, the $\chi(F)$ is undefined. We can treat it as a special case and assume harmonicity of 0 for all incommensurable sets. The Formula 4.3 can be rewritten by replacing $|H(F)|$ with $\frac{\max(F)}{\gcd(F)}$ according to Formula 4.2 and writing $\chi(F)$ so that there are two sets as inputs, like in Formula 4.1 for the affinity. This way, we get the following formula for harmonicity:

$$\chi(F, F') = \begin{cases} \gcd(F \cup F') \dfrac{|F \cup F'|}{\max(F \cup F')}, & \exists \gcd(F \cup F') \\ 0, & \nexists \gcd(F \cup F') \end{cases} \tag{4.4}$$

This formula accounts for both commensurable and incommensurable sets, which is relevant for mathematical strictness but is not relevant for real-world sounds, as any physical frequency set is commensurable. Just like affinity $\alpha$, harmonicity $\chi$ is a number between 0 and 1, where 0 means absolute lack of harmonicity and 1 means that $F$ and $F'$ form a single pitch.

The idea behind the harmonicity measure $\chi$ may be categorized as pattern-matching theory according to [1]. It is identical to the measure of harmonicity from the works [6] and [7] but reformulated in set-theoretic language. The work [6] focuses on rank ordering scales by their harmonicity, and the work [7] ranks chords and intervals by the same measure. Both studies show that harmonicity, defined in this way, is a strong predictor of consonance preferences. In this work, I view harmonicity as an internal, implicit characteristic of an arbitrary set of frequencies. It can be applied to inharmonic sounds, intervals, or chords and used to determine dynamic tuning on the fly rather than ranking static tunings.

To sum up. Harmonicity is the measure of how well individual pitches in the set form a larger single pitch, which is defined as a harmonic set. If individual pitches relate to each other as irrational numbers, they are completely disjoint and do not form a larger whole. If their relationship is rational, then they either form a single pitch or a subset of it. The value of harmonicity $\chi$ measures how close a set of pitches is to a single pitch. Sounds with a high degree of harmonicity are perceived as smoother and more stable, which satisfies our expectation for a correct measure of consonance. The reason for this is shown in Figure 3.5: harmonic sounds have more regular, periodic waveforms than other sounds.

One aspect that could be improved is the definition of what constitutes a single pitch. We consider only sets that satisfy Formula 3.4 to be single pitches. Thus, removing one or several arbitrary partials from $f_F \mathbb{N}_k$ will lower the harmonicity. That makes common timbres, such as a square wave or high-passed sounds, have an intrinsic degree of dissonance. That is counterintuitive and perhaps is too strict. However, I do not have a better suggestion for this measure of consonance than the one defined by Formula 4.4.

There may be many ways to quantify consonance using a set-theoretic approach. The ones used in the article are the simplest ones that are closely related to the prominent empirical theories. I admit that there is plenty of room for improvement, as many properties of elementary oscillations were omitted, such as amplitude and phase information. There may be a significant impact on consonance metrics from those additional parameters. I welcome anyone interested in participating to improve the theory. My goal is to have a simple, practically applicable theory that can lead to the creation of new instruments and music.

# 5 From consonance to tuning

In Chapter 1.2, we set out to find a natural solution to the tuning problem. That solution is a tuning system that arises naturally from mathematical objects and is independent of human values or perceptions, yet in harmony with them. We postulated a definition of consonance and suggested two quantities: affinity $\alpha$ and harmonicity $\chi$ that could serve as measures of consonance. Using them allows us to determine a set of intervals at which any two arbitrary sounds harmonize with each other, thereby providing a natural tuning system. Individually, affinity and harmonicity generate different sets of intervals, thus providing a very rich framework for tuning derivation. That flexibility can serve as an artistic tool to find new ways of expression. In the following chapters, I will propose the simplest ways to use these consonance measures to derive tuning systems and discuss their advantages and disadvantages.

## 5.1 Affinitive tuning system

At first, let's see what intervals are generated from the affinity. In Formula 4.1, affinity is a function of 2 arguments, which are sets $F$ and $F'$. Let's refer to $F$ as a **contextual set** and to $F'$ as a **complementary set.** Contextual set represents the frequencies of a note or a chord that serve as musical context for a complementary set. In other words, let's think of $F'$ as sounding relative to $F$, in the context of $F.$ As a real-life example, we can think of a choir where $F$ is a drone note that creates context for a melody. A note in a melody is represented by $F'$. Without context, we cannot speak of tuning, as any pitch is allowed. Tuning is a relative phenomenon, not an absolute one.

Both $F$ and $F'$ are arbitrary sounds. If they share some frequencies, they will have a non-zero affinity $\alpha(F, F')$ according to Formula 4.1. However, if they don't, we can always transpose $F'$ on some interval $t$ so that $\alpha(F, tF') > 0$. Let's think of a choir example again. We have a drone note $F$ relative to which another singer is about to take a note $F'$. At first, the singer misses the note, so that $\alpha(F, F') = 0$, which sounds out of tune. But quickly he or she corrects it by shifting $F'$ on an interval $t$, so that the frequencies of both notes align and $\alpha(F, tF') > 0$. Depending on the structure of $F$ and $F'$ there can be many such intervals with non-zero affinity. The set of all such intervals we can call affinitive intervals $\mathcal{A}(F, F')$ and define as:

$$\mathcal{A}(F, F') = \{t \mid \alpha(F, tF') > 0\} \tag{5.1}$$

Both $F'$ and $F$ are arbitrary sets that can represent a single note or a chord, be harmonic or inharmonic. The difference is that set $F$ represents a static, contextual sound, and $F'$ is the sound that we want to add relative to $F$. The set of affinitive intervals $\mathcal{A}(F, F')$ is all the intervals we can transpose $F'$ so it fits to $F$.

It may seem that finding $\mathcal{A}(F, F')$ is a difficult task, but it's actually quite trivial. Let's remember that according to Formula 4.1 the affinity of two sets is $0$ when their intersection is an empty set. Thus, we can rewrite 5.1 to be:

$$\mathcal{A}(F, F') = \{t \mid F \cap tF' \neq \emptyset\} \tag{5.2}$$

It is evident that to get a non-empty intersection, a number $t$ must be such that upon multiplication on an element of set $F'$ it should produce an element of $F$, or $tf' = f$ where $f \in F$ and $f' \in F'$. Solving this simple equation for $t$ gives us the following set $\mathcal{A}(F, F') = \left\{\frac{f}{f'} \mid f \in F, f' \in F'\right\}$. In simple words, a set of affinitive intervals is a set of ratios of every frequency in the contextual set $F$ on every frequency of the complementary set $F'$, which according to notation in Table 8.2 can be written as:

$$\mathcal{A}(F, F') = \frac{F}{F'} \tag{5.3}$$

Let's see how that can be applied in practice. For simplicity, let's assume that both sets are simple harmonic sets with 6 partials that represent the note $C4$ with a fundamental of 262Hz.

$$F = F' = C4_6 = 262 \cdot \mathbb{N}_6 \tag{5.4}$$

According to Formula 5.3, affinitive intervals are the ratios between every element of $F$ and $F'$. The easy way to find them is to build a table, like Table 5.1, where each cell is a simplified ratio of an element of $F$ (top row) to an element $F'$ (first column).

Table 5.1 - Derivation of affinitive intervals for the sets $F$ and $F'$ defined by Formula 5.4.

| $F'$ (Hz) \ $F$ (Hz) | 262 | 524 | 786 | 1048 | 1310 | 1572 |
|---|---|---|---|---|---|---|
| **262** | $\frac{1}{1}$ | $\frac{2}{1}$ | $\frac{3}{1}$ | $\frac{4}{1}$ | $\frac{5}{1}$ | $\frac{6}{1}$ |
| **524** | $\frac{1}{2}$ | $\frac{1}{1}$ | $\frac{3}{2}$ | $\frac{2}{1}$ | $\frac{5}{2}$ | $\frac{3}{1}$ |
| **786** | $\frac{1}{3}$ | $\frac{2}{3}$ | $\frac{1}{1}$ | $\frac{4}{3}$ | $\frac{5}{3}$ | $\frac{2}{1}$ |
| **1048** | $\frac{1}{4}$ | $\frac{1}{2}$ | $\frac{3}{4}$ | $\frac{1}{1}$ | $\frac{5}{4}$ | $\frac{3}{2}$ |
| **1310** | $\frac{1}{5}$ | $\frac{2}{5}$ | $\frac{3}{5}$ | $\frac{4}{5}$ | $\frac{1}{1}$ | $\frac{6}{5}$ |
| **1572** | $\frac{1}{6}$ | $\frac{1}{3}$ | $\frac{1}{2}$ | $\frac{2}{3}$ | $\frac{5}{6}$ | $\frac{1}{1}$ |

As you can see from the table, there are a total of 36 intervals, but many of them repeat more than once. Because sets do not contain duplicate elements, the actual number of unique elements is 23, and the set looks like this:

$$\mathcal{A}(C4_6, C4_6) = \left\{\frac{1}{6}, \frac{1}{5}, \frac{1}{4}, \frac{1}{3}, \frac{2}{5}, \frac{1}{2}, \frac{3}{5}, \frac{2}{3}, \frac{3}{4}, \frac{4}{5}, \frac{5}{6}, 1, \frac{6}{5}, \frac{5}{4}, \frac{4}{3}, \frac{3}{2}, \frac{5}{3}, 2, \frac{5}{2}, 3, 4, 5, 6\right\} \qquad 5.5$$

This is a complete list of intervals for which $C4_6$will harmonize with $C4_6$. The frequency structure of the note naturally elevates just 23 numbers from the infinite list of possibilities. Changing the contextual or complementary sets will elevate other intervals in a mathematically strict manner.

Traditionally, having a set of affinitive intervals would be considered as having an affinitive tuning system, but the set-theoretic approach can be pushed further. Each interval can be paired with a corresponding value of consonance. This way, the tuning system becomes a hierarchical system of intervals, from most to least consonant. In addition to calculating a value for affinity for each interval, nothing stops us from accounting for harmonicity as well. We can do it by finding the arithmetic mean of affinity and harmonicity for a given interval. We will refer to that value as **total consonance,** which is defined by the following formula:

$$\sigma(F,F') = \frac{\alpha(F,F') + \chi(F,F')}{2} \tag{5.6}$$

The arithmetic mean is the simplest way to combine both contributions that yields a value for total consonance between 0 and 1, where 1 indicates a perfect consonance between $F$ and $F'$ and 0 a complete lack of consonance. Now we can build what we will call an affinitive tuning system $T_{\mathcal{A}}$ as a set of ordered pairs of affinitive intervals and their corresponding values of total consonance:

$$T_{\mathcal{A}}(F,F') = \left\{\langle t, \sigma(F, tF')\rangle \mid t \in \frac{F}{F'}\right\} \tag{5.7}$$

The tuning $T_{\mathcal{A}}(C4_6, C4_6)$ that corresponds to the set of intervals in Formula 5.5 is depicted in Figure 5.1. On the horizontal axis, we plot the value for the interval $t$ in cents, and on the vertical axis, the value of total consonance $\sigma(C4_6, t \cdot C4_6)$. For convenience in determining the notes of the tuning, the piano keyboard is shown at the bottom of the chart. Each piano key corresponds to the frequency range for the notes in Table 8.3. The red key highlights the note of musical context (in this case, it is $C4$). One can see that each interval from Formula 5.5 is present in the graph and has a non-zero total consonance. The value of total consonance for the interval $t$ corresponds to how consonant both $F$ and $tF'$ will be if sounding together. For example, let's look at the interval of the perfect fifth with $t = 3/2$ or $702$ cents. Transposing the complementary set $F' = C4_6$ on that interval yields the note $G4_6 = \frac{3}{2}C4_6$ and using Formula 5.6 gives $\sigma(C4_6, G4_6) = 0.44$. That means that if you add $G4_6$ to $C4_6$ the consonance will be $0.44$. On the graph, the value for total consonance is shown as a sum of the contributions of harmonicity (orange lines) and affinity (green lines) stacked on top of each other. So, the value $0.44$ for the perfect fifth is the total length of the vertical line (orange + green), and it is just shy of $0.5$.

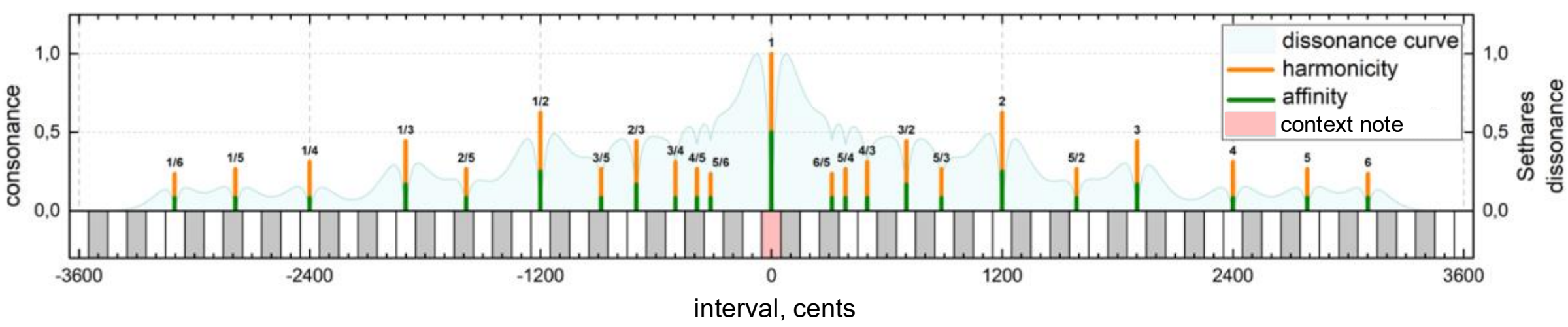

Figure 5.1 – Affinitive tuning $T_{\mathcal{A}}(C4_6, C4_6)$, where $C4_6 = 262 \cdot \mathbb{N}_6$. Black numbers represent the exact numeric value of the interval $t$, orange and green bars show the values of harmonicity and affinity as parts of the total consonance $\sigma$. The blue curve is a Sethares dissonance curve for $C4_6$ the spectrum. Note that sharp minima in the dissonance curve mark the same intervals as in the affinitive tuning. Below the graph, the equally tempered piano keyboard is presented as a reference, with the red note marking the note of a contextual set $F$.

We see that each interval of affinitive tuning coincides with a sharp minimum of the dissonance curve (blue curve) calculated for the $C4_6$ spectrum. That is expected, as those minima of the dissonance curve appear when the frequencies of two sounds coincide, which is exactly the intervals at which affinity is greater than zero. The consonance values show which interval is most consonant relative to the musical context. The only perfectly consonant interval with $\sigma = 1$ is unison 1:1, followed by octaves 2:1 and 1:2 with $\sigma = 0.625$, then perfect fifths 3:2 and 2:3 with $\sigma = 0.44$, and other intervals having lower values of $\sigma$.

The peculiar thing is that the tuning is symmetrical around the unison. That symmetry does not mean that the notes to the left of unison are the same as those to the right, as one might assume. In fact, the interval $4/5$ marks $G\#3$, which is missing in the right half of the tuning. It is important to note that affinitive tuning does not suppose nor predict the octave equivalence[30]. That means that even though $G\#3$ is part of the tuning, $G\#4$ is not. If we include octave equivalence as an additional axiom, we can transpose all the intervals into a single octave (Figure 5.2). Notice that after transposition, the interval $4/5$ becomes $8/5$ with zero value of affinity.

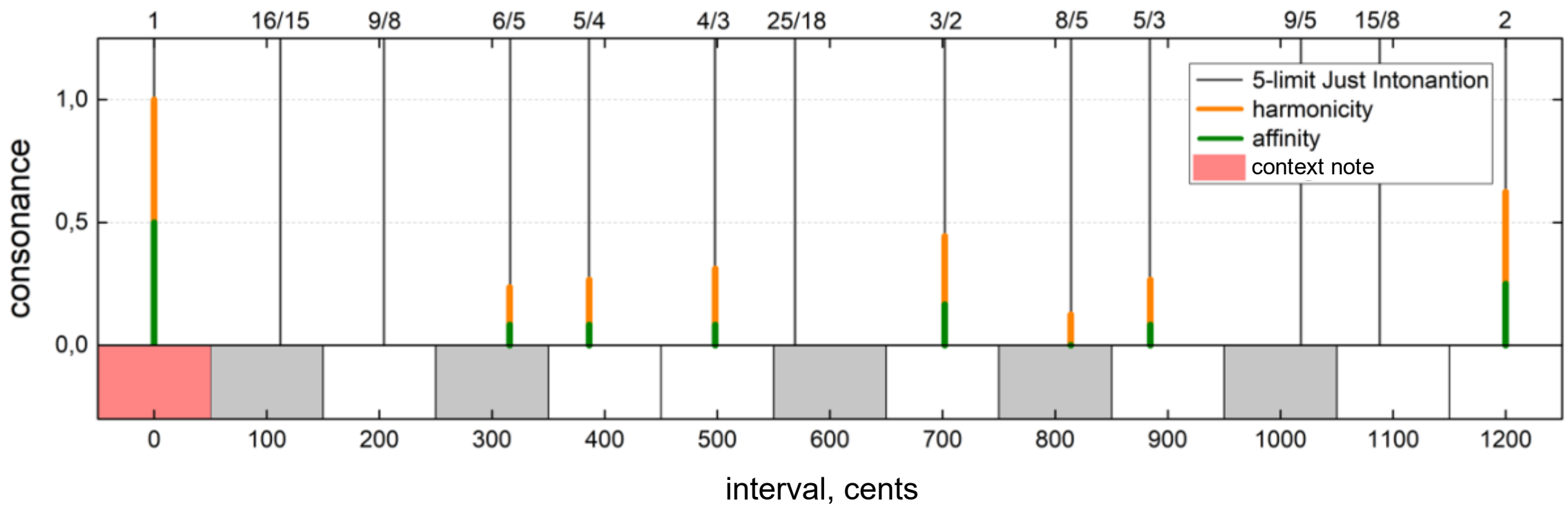

Figure 5.2 – The same affinitive tuning $T_{\mathcal{A}}(C4_6, C4_6)$ as in the previous figure, but where all of the intervals are transposed to be within a range of a single octave. Black vertical lines mark the intervals of 5-limit[31] Just Intonation for 12 notes of the chromatic scale. 5-limit intervals are chosen to be the simplest ratios that best approximate the corresponding 7-limit intervals.

Comparing this tuning with 5-limit JI, we see that JI includes affinitive tuning as a subset. The significant differences between JI and affinitive tuning are the way a practitioner arrives at a particular set of intervals. In JI, there are many ways to construct a set of intervals. For example, instead of using $25/18$ for the augmented fourth, like in Figure 5.2, it is common to use the ratio $45/32$. In JI, it is up to a practitioner to pick a set of intervals from all rational numbers. In affinitive tuning, the practitioner chooses sets $F$ and $F'$, but the tuning is the result of a strict mathematical procedure.

[30] Octave equivalence refers to the observation that notes that are octave apart can be considered to be the same note.
[31] The limit of JI refers to the maximum prime number that can be used to construct ratios of intervals. In 5-limit JI the ratios of intervals are products of prime numbers up to 5, in 7-limit JI up to 7. For more details refer to [8].

Affinitive tuning is the product of two sets, $F$ and $F'$. Any change in one of them will be reflected in the tuning. Let's investigate how affinitive tuning changes if the contextual set is an interval or a chord rather than a single note. We see in Figure 5.3 that with every additional note in the contextual set $F$, the number of intervals in the tuning increases, and the consonance values for each interval change. That shows the dynamic nature of tuning. As the musical piece progresses, the contextual set changes, reflecting shifts in harmony, and the affinitive tuning changes with it.

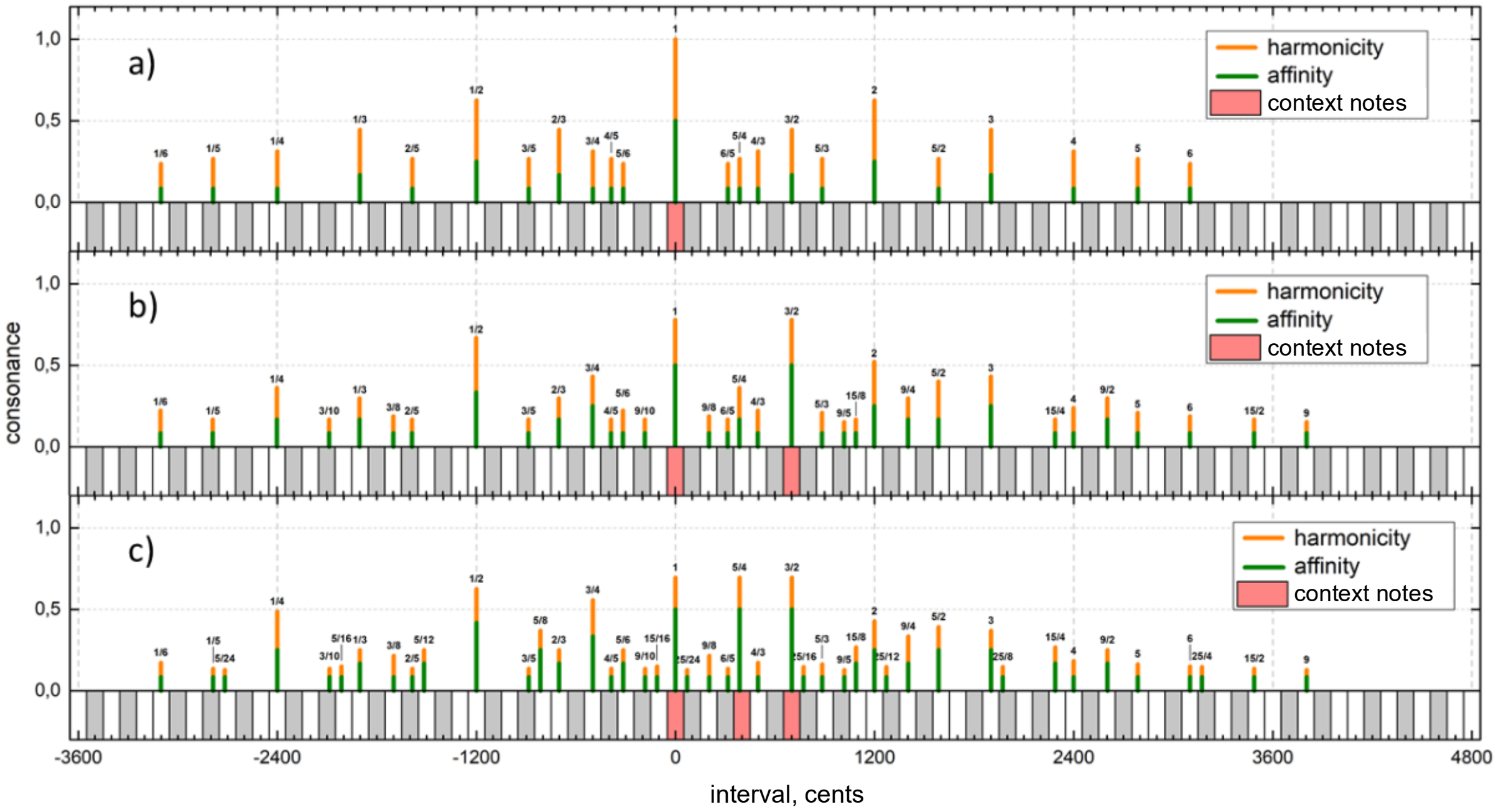

Figure 5.3 – Demonstration of the dynamic nature of the affinitive tuning with the change of the contextual set $F$. In a) set $F = C4_6 = 262 \cdot \mathbb{N}_6$. In b) $F = C4_6 \cup G4_6$ where $G4_6 = \frac{3}{2} C4_6$. In c) $F = C4_6 \cup E4_6 \cup G4_6$ where $E4_6 = \frac{5}{4} C4_6$.

The maximum value of consonance reflects the consonance of the current musical context. The more complex the harmony, the lower the consonance value. The difference in the dynamism of affinitive tuning from other dynamic tuning systems is that the frequencies in the contextual set $F$ do not change as music progresses. Many other dynamic tuning systems, as described in [4], aim to fix the 12-tet tuning while preserving the 12-tet keyboard interface.[32] Such tunings exhibit a class of artifacts in which notes in the melody affect the pitch of notes in harmony, causing them to move up and down to minimize dissonance. Such a situation does not occur in the affinitive tuning system; all frequencies in the context are fixed.

[32] The examples of such dynamic tunings would be a Sethares dissonance curve based tuning or a Hermode tuning. Good review is found in [4]. Also, an interesting continuation of the Sethares work is Dynamic Tonality [13].

A similar dynamic behavior emerges in Sethares' theory and in free JI. The benefit of the affinitive tuning is very low computational requirements and the ability to quantify consonance of intervals using both affinity and harmonicity measures of consonance. Accounting for harmonicity does privilege harmonic sounds over inharmonic, which is evident from Figure 5.4 which shows the affinitive tuning for highly inharmonic sound $F_{inharmonic}$

$$F_{inharmonic} = 262 \cdot \{1, 2.76, 5.41, 8.94, 13.35, 18.65\} \qquad 5.8$$

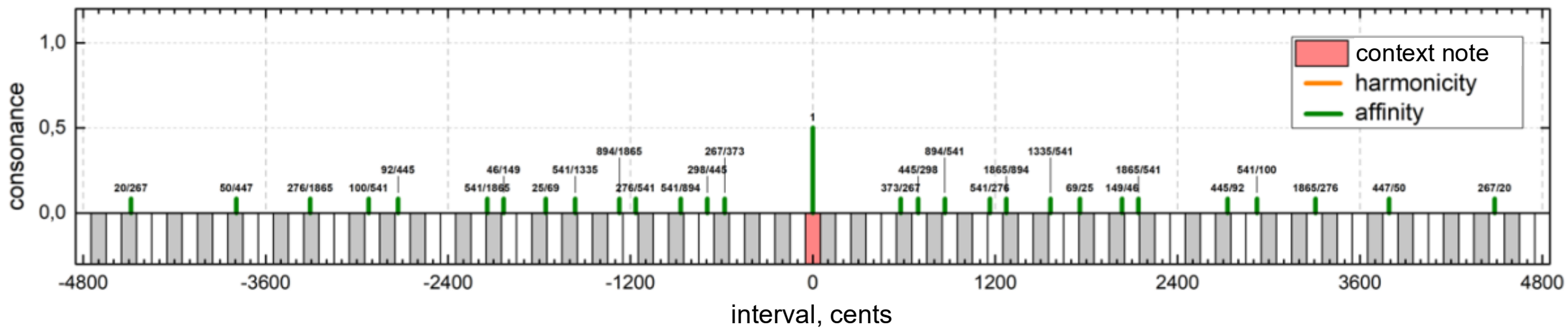

Figure 5.4 – Affinitive tuning for the inharmonic sounds $T_{\mathcal{A}}(F_{inharmonic}, F_{inharmonic})$. A very high degree of inharmonicity of sets reduces the harmonicity contribution to near-zero values, which makes affinity a dominant contribution to consonance.

One can see that the harmonicity contribution is practically non-existent, as it is orders of magnitude lower than affinity. In Sethares' framework, there is no differentiation or ranking between different timbres. All of them are equally usable and depend on the practitioner's taste. In the suggested framework, there is a visible differentiation between timbres and tunings. Harmonic sounds, as in Figure 5.1, seem to produce much richer and more expressive tuning than inharmonic sounds in Figure 5.4. That is evident from a greater variance of consonance values between different intervals. Other inharmonic timbres can be more expressive than $F_{inharmonic}$, but with a set-theoretic approach, we have an objective measure for judging them.

In the case of affinitive tuning, harmonicity plays an important role, but it does not affect the set of intervals in tuning. It is possible that some meaningful intervals with large consonance values may be missed out if we only use affinity to determine intervals. The most glaring issue that indicates that we have to find a way to include harmonicity into interval derivation is that if sets $F$ and $F'$ have only one frequency. In this case, the tuning will consist of only a single interval – a unison. Indeed, as we will see from the next chapter, affinity may miss a lot of intervals with large values of consonance. Thus, let's investigate which tuning emerges from harmonicity.

## 5.2 Harmonic tuning system

Just like affinity, harmonicity is a function of the contextual set $F$ and the complementary set $F'$. To get the affinitive tuning, we came up with a procedure of finding all intervals $t$ for which the value of affinity $\alpha(F, tF')$ is greater than zero (Formula 5.1). Such an approach generated a tuning with a moderate number of intervals, which is important as having too few intervals is not very expressive, and too many are overwhelming. Unfortunately, the same approach does not work as well for harmonicity. The reason for that is that harmonicity is strictly zero when a set of frequencies is incommensurable (see Chapter 4.3). That means that if one of the sets $F$ and $F'$ is incommensurable, the harmonicity will be 0 for all intervals $t$. If, on the other hand, both sets $F$ and $F'$ are commensurable, then harmonicity will be 0 for all irrational $t$ and greater than zero for all rational $t$.

The same conclusion, that every rational number is to some degree consonant, was reached by theoreticians of JI. The degree of consonance in JI is measured as the simplicity of the ratio that constitutes that interval [8]. Let's investigate the distribution of harmonicity in the simplest case when both $F$ and $F'$ have a single element and equal to $C4_1 = 262 \cdot \mathbb{N}_1$. Since we cannot check all rational numbers, we have to set arbitrary limits. Let's calculate a total consonance for all rational numbers with a denominator less than $60$ and in the span of $\pm 3$ octaves (Figure 5.5).

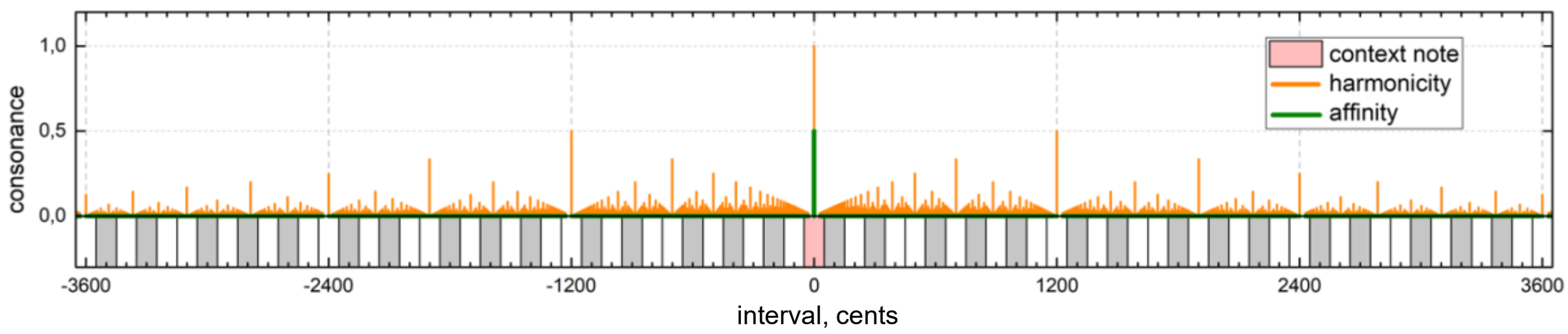

Figure 5.5 - Distribution of harmonicity for a span of intervals ranging from -3 to +3 octaves. The intervals are chosen so that the denominator has a maximum value of 60. Consonance is calculated for $F = F' = 262 \cdot \mathbb{N}_1$ being a single partial with a frequency that corresponds to the note $C4$. Due to the high density of intervals, the labels for their ratios are omitted.

As seen from the graph, each value of harmonicity is a local maximum surrounded by much smaller values. That means that tiny deviations from precise ratios cause a dramatic reduction of harmonicity.[33]

[33] Floating point arithmetic proved to be unsuitable for plotting a harmonicity distribution. Tiniest deviations from the exact ratios, such as $4/3$, caused a near zero value of harmonicity. Thus, a rational number arithmetic using Fraction.js library [14] had to be used to derive values of intervals.

The graph in Figure 5.5 demonstrates an issue with affinitive tuning that was stated at the end of the previous chapter. For sets with a single partial like $C4_1$, only the unison 1:1 has a non-zero value of the affinity, while there are plenty of intervals with large values of harmonicity to pick from. What is remarkable is the beautiful fractal structure of consonance that can be described by a modified Thomae's function[34]:

$$\tau_m(t) = \begin{cases} \frac{1}{\max(p,q)} & if\ t = \frac{p}{q}\ (t\ is\ rational), with\ p \in \mathbb{Z}\ and\ q \in \mathbb{N}\ coprime \\ 0 & if\ t\ is\ irrational \end{cases} \tag{5.9}$$

The Figure 5.6 demonstrates that graph of Formula 5.9 (blue lines) coincides with the distribution of total consonance in Figure 5.5 (orange lines in Figure 5.6), meaning that it is the same function.

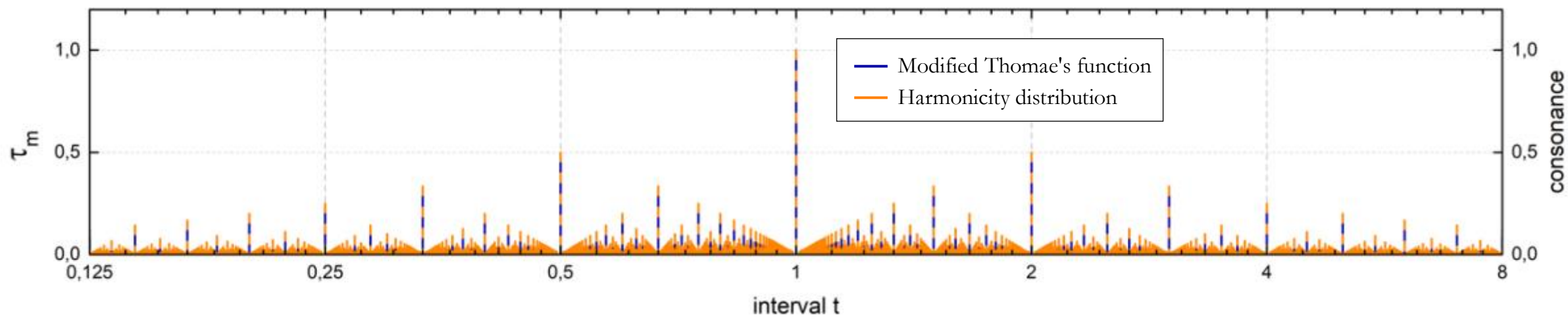

Figure 5.6 – Comparison of the distribution of total consonance for $F = F' = C4_1 = 262 \cdot \mathbb{N}_1$ (orange lines) on the right axis with modified Thomae's function described by equation 5.9 (blue line) on the left axis.

This non-trivial result can be proven algebraically using the definitions of harmonicity and affinity, which is done in Chapter 8.5.

The Formula 5.9 proves the definition of consonance used by JI theoreticians, that the consonance of an interval is inversely proportional to the complexity of its ratio. Complexity of the ratio in Formula 5.9 is represented by the value $\max(p,q)$. Thus, the interval is more complex the larger the number needed to describe it. This way, the perfect fifth 3:2 is more consonant than the major third 5:4, and the direction of the interval does not influence the consonance rating[35]. Also, notice that this result is valid for the total consonance that includes affinity. The distribution of harmonicity does not follow Formula 5.9, as it has the same values for octave and unison as seen in Figure 5.7.

[34] See Chapter 8.4 for more info on Thomae's function.
[35] For example, the perfect fifth built up from C4 to G4 is represented by the interval 3:2, while if built down from C4 to F3 is represented by 2:3. Both are perfect fifth and both will have the same complexity $\max(3,2) = \max(2,3)$.

This result holds only for sounds with a single partial. More complex sounds alter the distribution of harmonicity. Figure 5.7 compares the harmonicity distribution for harmonic sounds with 1, 6, and 256 partials. One can see a noticeable difference in harmonicity values between the 1- and 6-partial cases, but beyond 6 partials, the distribution changes little.

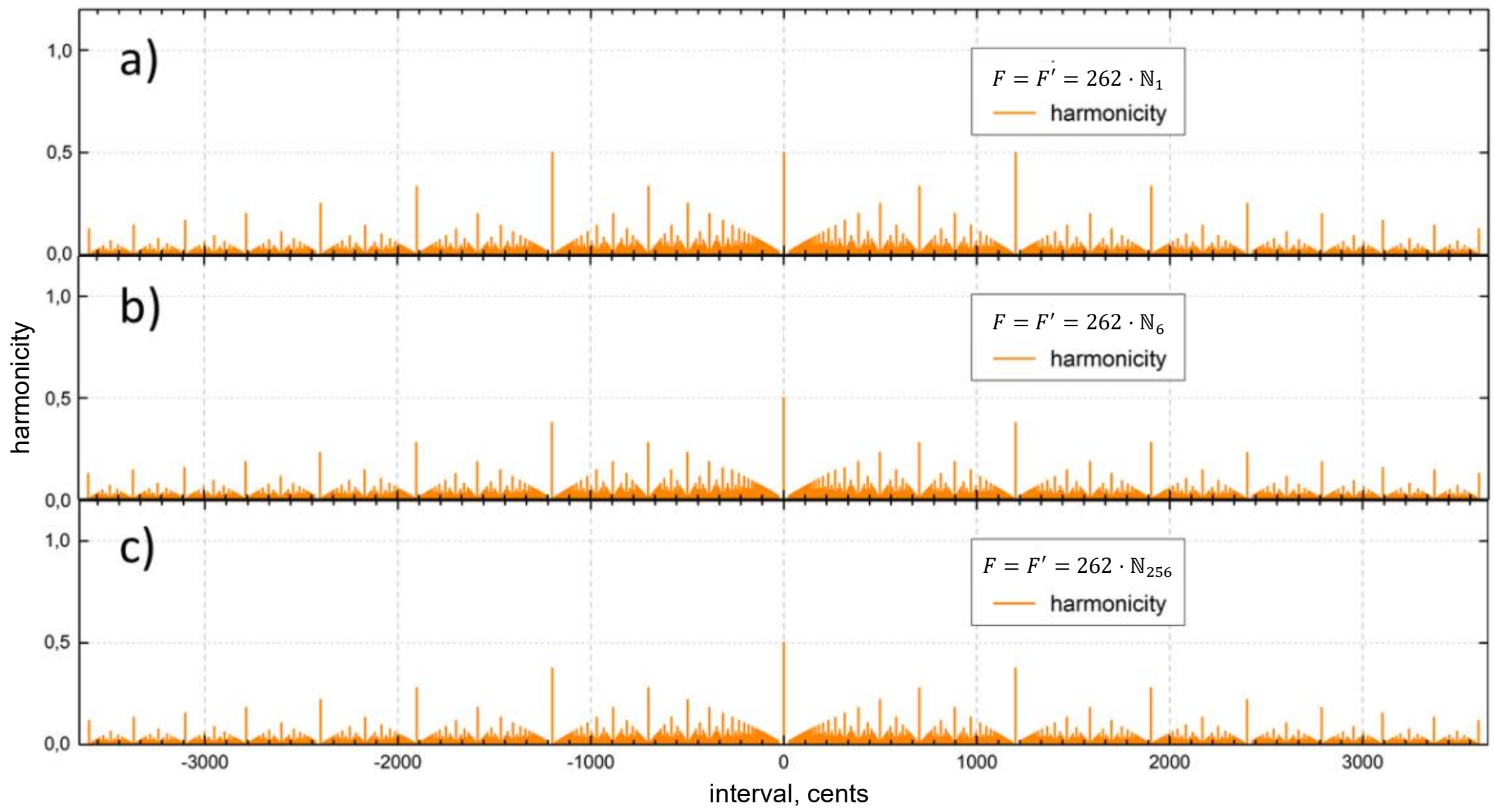

Figure 5.7 – The change in harmonicity distribution with increasing number of harmonics, affinity values are omitted. a) Harmonicity distribution for a single partial $F = F' = 262 \cdot \mathbb{N}_1$. b) Harmonicity distribution for harmonic sound with six partials $F = F' = 262 \cdot \mathbb{N}_6$. c) Harmonicity distribution for harmonic sound with 256 partials $F = F' = 262 \cdot \mathbb{N}_{256}$. Increasing the number of partials past six does not lead to a dramatic change in harmonicity distribution.

More noticeable changes to the harmonicity distribution happen when more notes are added to the context set. Figure 5.8 demonstrates what will happen with the distribution of harmonicity if we start adding additional notes to the context, starting with a single $C4_6 = 262 \cdot \mathbb{N}_6$.

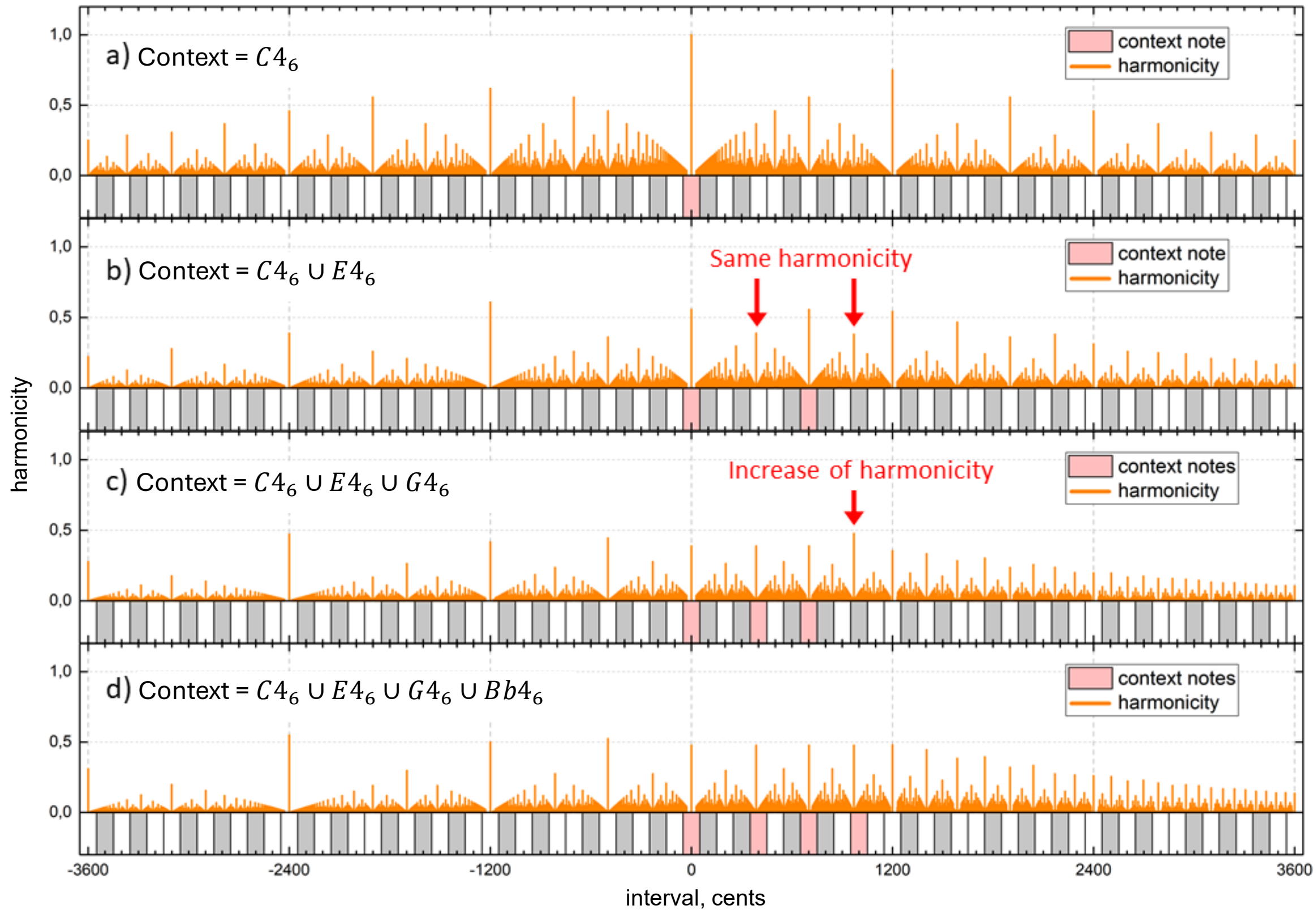

Figure 5.8 – Changes in harmonicity distribution caused by changes in the contextual set. Affinity values are omitted. a) The context is a single note $F = C4_6$. b) The context is an interval of a perfect fifth $F = C4_6 \cup G4_6$. Note the same values for harmonicity for major third 5:4 and harmonic seventh 7:4. c) The harmonicity distribution for a major triad $F = C4_6 \cup E4_6 \cup G4_6$. Note the increase in harmonicity for $Bb4_6$. d) The harmonicity distribution for $C7$ the chord $F = C4_6 \cup E4_6 \cup G4_6 \cup Bb4_6$.

When considering what note to add to a single $C4_6$, the most consonant interval (as measured by harmonicity alone) is a perfect fifth up, which gives $G4_6 = \frac{3}{2} \cdot C4_6$. Adding it to the context leads to the surprising result of notes $E4_6 = \frac{5}{4} \cdot C4_6$ and $B\flat4_6 = \frac{7}{4} \cdot C4_6$ having the same value of harmonicity. Adding $E4_6$ to the context to form a major triad further emphasizes the harmonic seventh $B\flat4_6$ as a note that increases the overall harmonicity of a chord! That runs contrary to the usual notion of a major triad being more consonant than a major seventh chord. One might take it as proof that a reliance on harmonicity as a sole measure of consonance is error-prone. As shown in Figure 5.9, adding affinity contribution changes consonance distribution in favor of common triadic harmony.

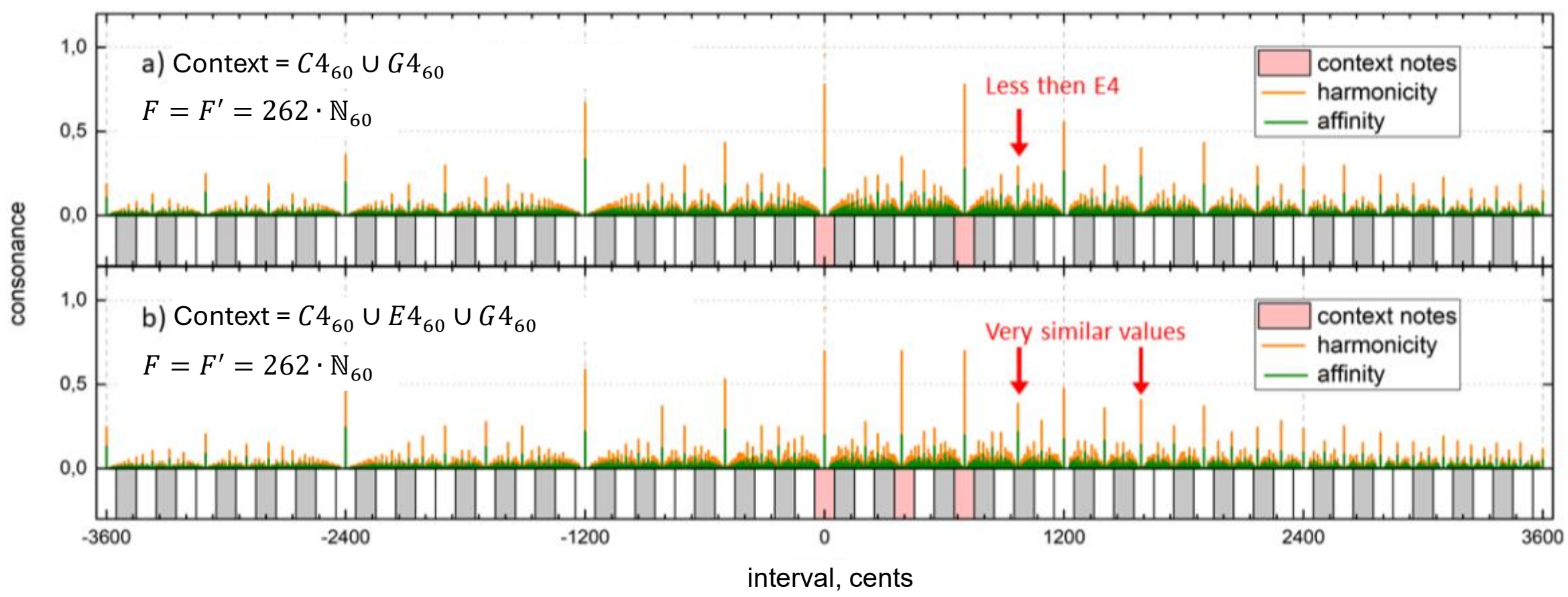

Figure 5.9 – Distribution of total consonance for a rich harmonic sound with 60 partials to demonstrate a preference for triadic harmony that happens when including affinity contribution. The increase in partial count is done to equalize contributions from harmonicity and affinity. A lower number of partials does not change the result. a) Harmonicity distribution for two notes in the context $F = C4_{60} \cup G4_{60}$, compared to Figure 5.8, the harmonic seventh does have a smaller value than the major third. b) Harmonicity distribution for the major triad in the context set. Harmonic seventh does not increase consonance as in Figure 5.8. However, the consonance value of $E5_{60}$ is the same as for the harmonic seventh $Bb4_{60}$.

As shown in the previous chapter, affinity becomes a dominant contribution to consonance when dealing with inharmonic sounds. Figure 5.10 shows the distribution of harmonicity for inharmonic sets, as in Formula 5.8. One can see that the overall shape is very similar to that of a single partial in Figure 5.7, but with harmonicity values several orders of magnitude smaller.

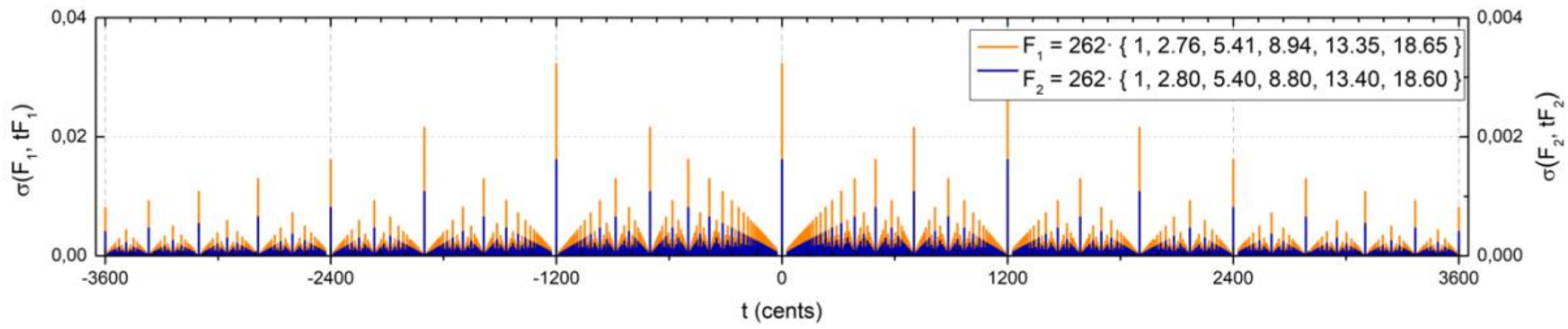

Figure 5.10 – Comparison of the harmonicity distribution of 2 inharmonic sounds. On the right axis - harmonicity for inharmonic sound $F_{inharmonic}$ described by Formula 5.8. On the left axis – harmonicity for the inharmonic sound made by rounding off the last digit of partials in $F_{inharmonic}$. Notice that rounding off a single digit increases harmonicity by almost 20 times.

As mentioned previously, harmonicity is very sensitive to the precision of the frequencies in a set. Rounding off a single digit can lead to a dramatic increase in the harmonicity of the sound, as shown in Figure 5.10.

Because harmonicity naturally prefers rational numbers as described by small ratios, one can say that JI is the natural harmonic tuning we were seeking to find in this chapter. That is true if we add octave equivalence as an additional axiom. Also, having strict measures of consonance provides more clarity to the consonance of various note combinations than in plain JI, and gives us a new way we can limit the number of intervals in the tuning. For example, we can only include the most consonant intervals that have a total consonance higher than some number $h$. This way, the set of harmonic intervals will be determined by the following formula:

$$\mathcal{H}(F, F', h) = \{t \mid \chi(F \cup tF') > h\} \quad 5.10$$

That lets us define the harmonic tuning in the following way:

$$T_{\mathcal{H}}(F, F', h) = \{\langle t, \sigma(F, tF')\rangle \mid t \in \mathcal{H}(F, F', h)\} \quad 5.11$$

Figure 5.11 shows that harmonic tuning is better suited for a sound with few partials than the affinitive tuning, given a carefully picked value of $h$. Having $h$ too small, generates too many intervals in Figure 5.11.b, but raising it to $h = 0.23$ filters out most of the intervals, leaving only the most consonant ones (Figure 5.11.c).

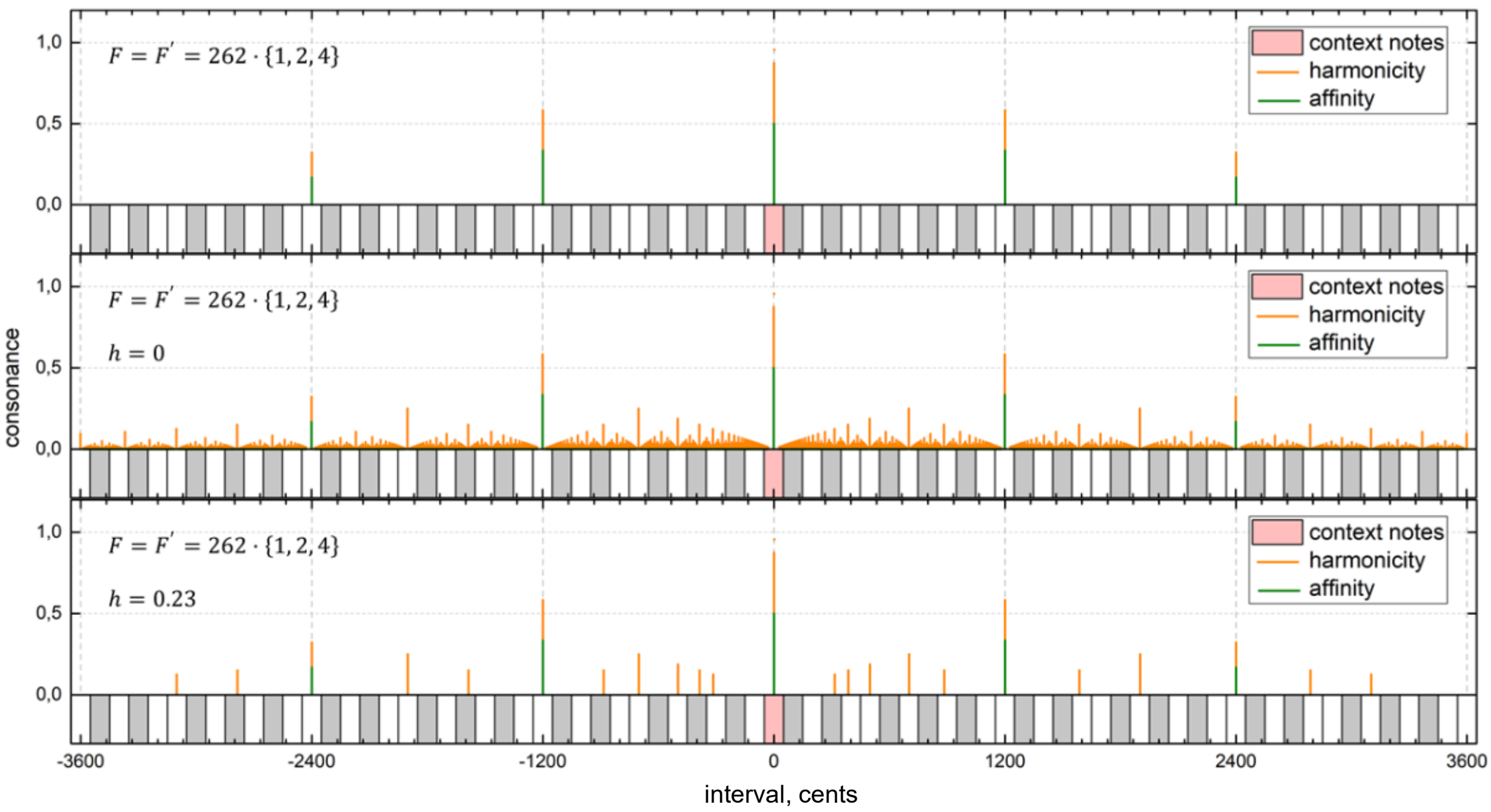

Figure 5.11 – Advantage of harmonic tuning for sounds with a small number of partials. a) Affinitive tuning for $F = F' = 262 \cdot \{1, 2, 4\}$. Tuning is useless for music-making, as it only includes octaves and double-octave intervals. b) Harmonic tuning with $h = 0$ for the same set as in a). It has too many intervals due to the low value of $h$. c) Harmonic tuning with $h = 0.23$ for the same sound as in a) and b). This value of $h$ filters out most of the intervals, leaving us with perfectly usable tuning.

Defined in this way, harmonic tuning suits sounds with few partials better than affinitive tuning. That advantage, however, comes with the much larger computational overhead, as we need to process hundreds or thousands of intervals for their consonance values to be able to pick the most consonant ones. Another drawback is the difficulty of using harmonic tuning for inharmonic sounds, as some intervals with non-zero affinity can be missed due to insufficient precision (Figure 5.14.a). The last issue is a very high sensitivity to deviation from simple ratios, which contradicts empirical observations. I believe it is possible to account for the human auditory system in various ways to alleviate that problem; however, doing that is not in the scope of this article.

Just as with affinitive tuning, harmonic tuning has its own set of advantages and disadvantages. The question is whether there is an alternative way to include harmonicity in tuning derivation?

## 5.3 Harmonic superset tuning system

In Chapter 4.3, we discussed how different frequencies of a harmonic set form a single pitch. We also discussed that any commensurable set $F$ has a harmonic superset $H(F)$, which means that $F$ is a part of a single pitch $H(F)$. That harmonic superset is a virtual pitch of that sound. In that sense, we may say that two notes comprising the major third interval in Figure 4.4 are two separate pitches, but at the same time, we can say that together they form a part of a single 88Hz pitch. This way, instead of finding affinitive intervals for the sets $F$ and $F'$, we can find them for the corresponding harmonic supersets $H(F)$ and $H(F')$.

Using supersets to generate intervals should provide an improvement for harmonic sounds with missing partials, like in Figure 5.11.a. The superset will fill the gaps in the spectrum and provide more intervals for tuning. However, if the sound consists of only a single frequency, the harmonic superset, as defined by Formula 4.2, will also have only one frequency. That can be fixed, as the condition $k = \frac{\max(F)}{\text{gcd}(F)}$ that limits the number of frequencies in the harmonic superset is the minimal condition that guarantees that all frequencies in $F$ are included in $H(F)$. Thus, if $k$ is too small to get a meaningful tuning, we can increase it by a number $n$ to get more intervals:

$$H_n(F) = \text{gcd}(F)\, \mathbb{N}_{k+n}, \qquad k = \frac{\max(F)}{\text{gcd}(F)} \qquad 5.12$$

To define the harmonic superset tuning system, we need to assign the values of consonance to each interval. We can measure consonance for either harmonic supersets $H_n(F)$ and $H_m(F')$ or the original sets $F$ and $F'$. I would argue that since frequencies in supersets are not present in the physical sound, they should not be treated equally to the real frequencies. Thus, the total consonance should be calculated for $F$ and $F'$, and not their supersets. Given that, we can define a harmonic superset tuning system with the following formula:

$$T_S(F, F', n, m) = \left\{\langle t, \sigma(F, tF')\rangle \mid t \in \mathcal{A}\left(H_n(F), H_m(F')\right)\right\} \quad 5.13$$

Arguments $n$ and $m$ can be used to increase the number of partials in harmonic supersets to provide a tuning for sets with a single partial, as shown in Figure 5.12.

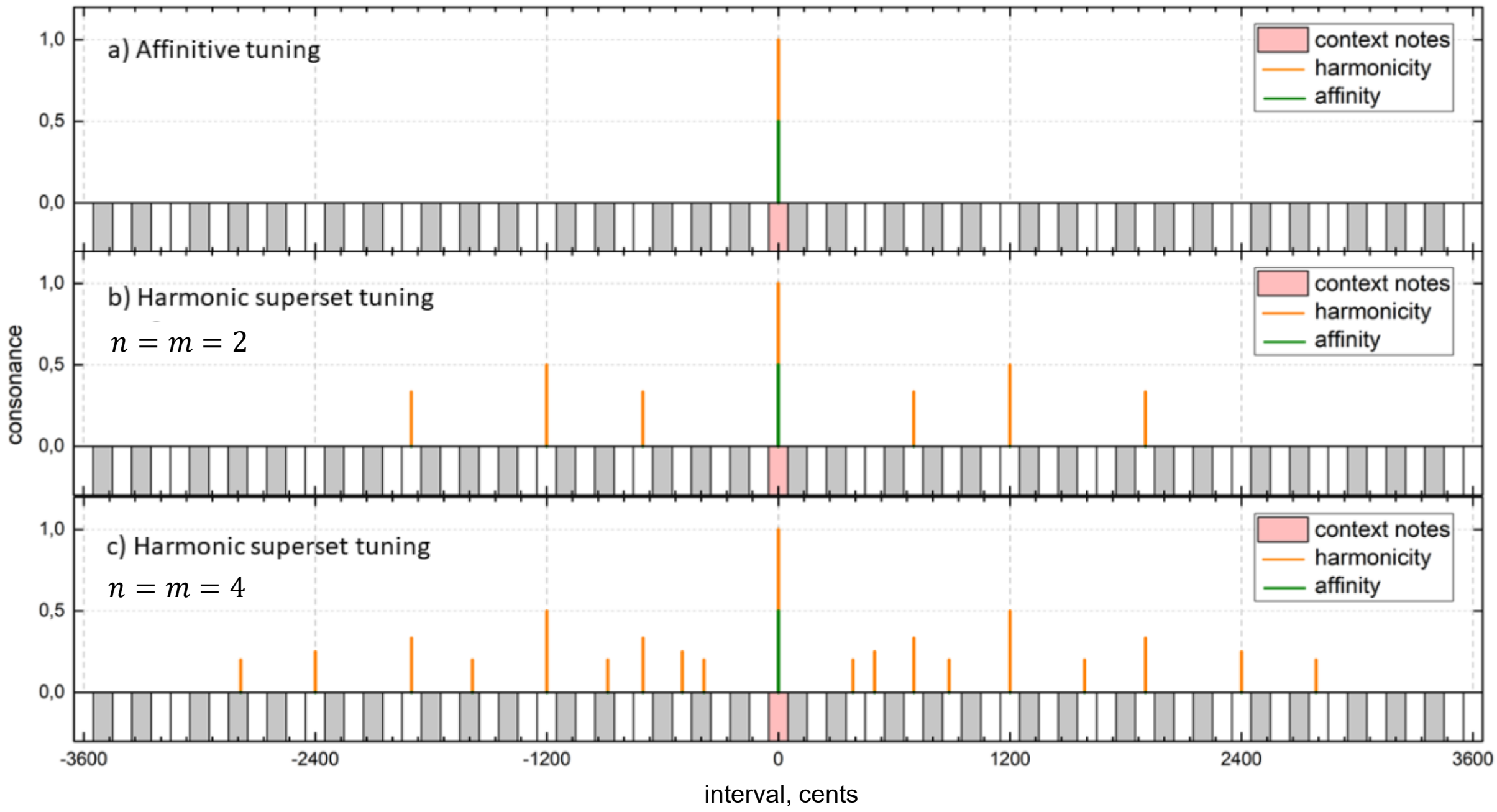

Figure 5.12 – Demonstration of the advantage of harmonic superset tuning system over the affinitive tuning system for the sounds with a single partial $F = F' = 262 \cdot \mathbb{N}_1$. a) Affinitive tuning for the sounds with a single frequency. b) Harmonic superset tuning with $n = m = 2$ and a few intervals. c) Harmonic superset tuning with $n = m = 4$ and a usable tuning that contains all of the most consonant intervals. Further increase of parameters $n$ and $m$ will lead to even richer sets of intervals.

Notice that most of the intervals (apart from unison 1:1) in Figure 5.12 have zero value for affinity. That is expected, as consonance is calculated for sets $F$ and $F'$, while intervals are generated from their supersets $H_n(F)$ and $H_m(F)$. Using more complex sounds will yield more intervals with non-zero affinity.

Affinitive tuning is a subset of harmonic superset tuning, because the condition for $k$ in Formula 5.12 guarantees that all partials in $F$ and $F'$ are included in supersets and thus, all the intervals they generate will be included in harmonic superset tuning. That is seen in Figure 5.13, where the change in the contextual set $F$ causes similar changes in the tuning as in the affinitive tuning in Figure 5.3. However, in contrast to the affinitive tuning, we see additional intervals with zero value of affinity in harmonic superset tuning.

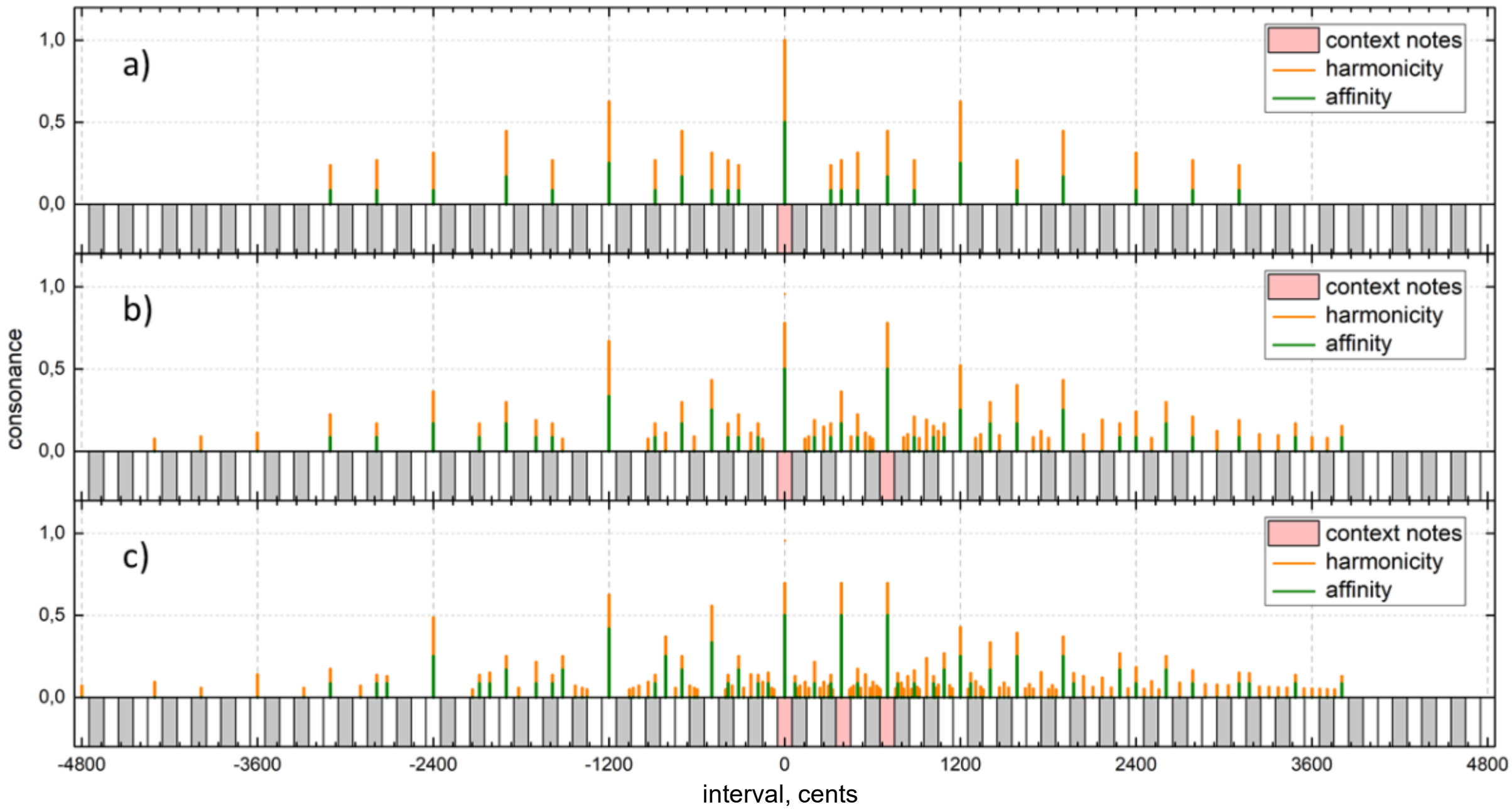

Figure 5.13 – Demonstration of the dynamic nature of harmonic superset tuning from the change of the context set. a) Harmonic superset tuning for the contextual set being a single note $F = C4_6 = 262 \cdot \mathbb{N}_6$. The tuning is identical to affinitive tuning in Figure 5.3.a b) Harmonic superset tuning for the interval of the perfect fifth $F = C4_6 \cup G4_6$. We notice additional intervals with zero affinity that are missing in affinitive tuning in Figure 5.3.b c) Harmonic superset tuning for the major triad $F = C4_6 \cup E4_6 \cup G4_6$. Comparison with Figure 5.3 reveals that affinitive tuning is a subset of harmonic superset tuning.

With the addition of the perfect fifth to the contextual set, we get an extra octave in the bass register compared to the affinitive tuning. That happens because with each addition of a new note, the virtual pitch of the contextual set $F$ drifts lower and lower, and the number of partials in $H_n(F)$ increases more and more to satisfy the minimal partial requirement. That leads to more intervals with zero affinity but non-zero harmonicity, resulting in a richer tuning.

Because harmonicity contribution is so sensitive to small deviations of partials (Figure 5.10), the affinitive tuning may be a better fit to inharmonic sounds as it does not spend extra computation resources on the intervals with dismissively small consonance values. However, compared to the harmonic tuning, there is no risk of missing valuable intervals, as harmonic superset tuning includes all intervals from affinitive tuning. The graphs in Figure 5.14 clearly demonstrate that advantage. Harmonic superset tuning does not depend on a precision parameter to capture all of the essential intervals from both affinitive and harmonic perspectives.

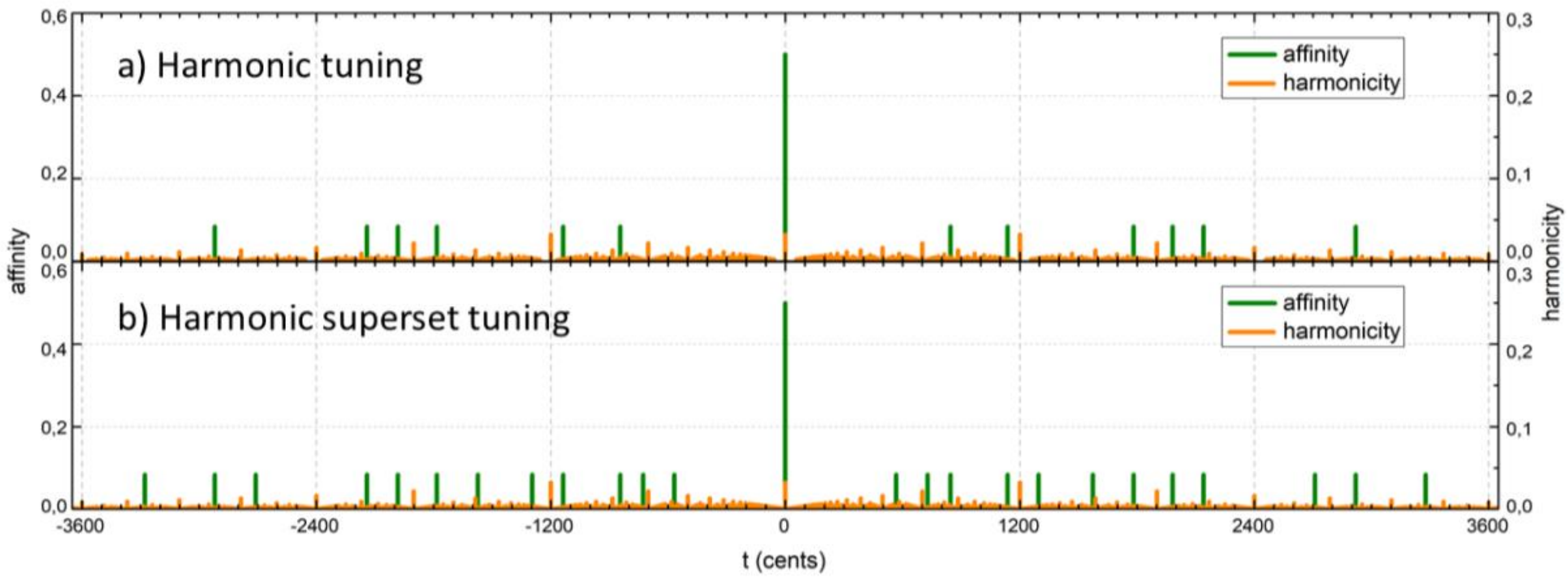

Figure 5.14 – Demonstration that harmonic superset tuning does not miss essential intervals for inharmonic sounds as does harmonic tuning. a) Harmonic tuning with a set of intervals limited by the maximum value of the denominator of 30 for the sets $F = F' = 262 \cdot \{1, 2.8, 5.4, 8.8, 13.4, 18.6\}$. b) Harmonic superset tuning for the same sets as a). Notice additional intervals with non-zero affinity that are missing in harmonic tuning in a). On both graphs, the values for harmonicity are scaled up twice to be visible and are plotted against the right axes.

Harmonic superset tuning improves upon the previous two systems by employing affinity and harmonicity in both aspects of the tuning system: generation of intervals and measurement of consonance. As a result, it addresses the drawbacks of other systems while not losing their benefits. It is a safe tuning system to pick that will not miss any important intervals and that allows the greatest flexibility for composition. Affinitive tuning is a good alternative to reduce the computational cost of deriving a tuning.

## 5.4 Discussion

In this chapter, I suggested three different tuning systems that naturally emerge from affinity and harmonicity consonance measures. Affinity generates affinitive tuning that is similar to the system based on dissonance curves suggested by Sethares in [4]. Harmonic tuning that emerges from harmonicity gives preference to simple-ratio intervals resembling JI [8]. Both set-theoretic systems improve upon the mentioned alternatives, as each interval is assigned a measure of total consonance. That is an important contribution, as it allows for the rank-ordering of intervals in a tuning by accounting for two of the most prevalent factors in the perception of consonance: spectral interference and harmonicity/periodicity.[36]

[36] According to [1], spectral interference and harmonicity/periodicity plays the largest role in the perception of consonance. Affinity, being a proxy measure for spectral interference, and harmonicity, a measure of harmonicity/periodicity, are both included in the total consonance that is defined by the Formula 5.6.

However, that improvement is not enough to check all the boxes in Table 2.1. The affinitive tuning may improve upon Sethares' tuning by reducing computation complexity and accounting for the spectrum's harmonicity, but, like Sethares' theory, it does not yield a usable tuning for sounds with few partials. A similar thing is true for harmonic tuning. Just like JI, it does not provide a good tuning for inharmonic timbres. Even though consonance rating may identify the intervals with a large value of affinity, it does not provide a guarantee that all such intervals will be found.

It is a nontrivial task to combine spectral interference and harmonicity/periodicity into a single framework for generating tuning. To the best of my knowledge, it has never been done to this day.[37] Luckily, the set-theoretic approach provides an almost trivial solution for that problem. By replacing original sets with corresponding harmonic supersets (corresponding virtual pitches), it is capable of generating tuning that is suitable for all types of sounds in Table 2.1. If true, that makes harmonic superset tuning the most broadly applicable tuning theory known to date.

On this note, it can be concluded that the aim of the paper is satisfied. The harmonic superset tuning system is a good candidate for a natural tuning system. It originates in pure mathematics and does not depend on subjective preferences. Potentially, its applicability is broader than any other tuning system that satisfies Ptolemy's condition for the validity of idealistic theory. However, further investigation is needed to support that claim. One way of doing that, is to test the system in practice and see what fruits come out of it.

The practical application necessarily raises the question of the appropriate interface to that tuning. We used a piano keyboard as a reference to show which interval corresponds to each note, but it does not suggest that using piano keyboard is in any sense appropriate. It is clear from Figure 5.14 that harmonic superset tuning is microtonal, and the density of intervals builds up quite quickly. It is possible to reduce the complexity to be compatible with the keyboard layout, but it may not be a good way forward. Like it was discussed in Chapter 2.3, the piano keyboard is a matched interface for 12-tet. Thus, moving away from 12-tet implies leaving the keyboard layout. I see the description and creation of such interface as my next endeavor, and I believe the digital technologies should make it accomplishable.

[37] In the work [1] authors claim to combine spectral interference and harmonicity into a single framework, however, the purpose of that framework was to explain the consonance preferences of people rather than providing a tuning system for musicians to use.

# 6 Conclusion

On that note, I would like to end the discussion of tuning systems and invite the reader to further investigate and experiment with a set-theoretic approach to musical tuning. We went through the history and philosophy surrounding the tuning theory and postulated a definition of consonance. We demonstrated how that definition can be captured by the mathematics of sets by defining two terms. The first one is affinity, which measures how well different notes fit together. The second one is harmonicity; it measures how the multiplicity of pitches forms a unity into a single pitch. Those two sides of consonance naturally give birth to affinitive and harmonic tuning systems, each with its own set of advantages and disadvantages. We also saw how the marriage of affinity and harmonicity produces the harmonic superset tuning system, which has the broadest range of applicability of any other theory and serves as a candidate for a natural tuning system. All the tuning systems we discovered are products of pure mathematics. Thus, their existence does not depend on the human auditory system or empirical observations.

I do not consider this work to be the final and comprehensive analysis of how set theory can be used to derive tuning systems. This is just my take on what can be done, a steppingstone to build upon. Can affinity and harmonicity be defined at a deeper level to strengthen the theoretical basis? I am sure they can. Can there be other aspects to consonance, sound, or tuning that I never thought of? Most certainly there are. Is it possible to build better tuning systems using the same definitions for consonance contributions? No doubt, it is. While writing this paper, I stumbled upon things I never expected to encounter. The unexpected links to other tuning theories, the simplicity of math, and the beauty of graphs - all prove to me that the set theoretic approach does reflect the nature of musical tuning in a deep sense. That inspires me to continue working on that topic by bringing that theory into practice. I don't know where that will lead me, but hopefully to some good music.

Follow my work on the New Tonality YouTube channel [9] and website [10]. A big thank you to all supporters on Ko-fi and all channel viewers. I express my gratitude to the authors of the cited papers and to everyone else who shared their love and knowledge of tuning theory and music. I am thankful to all my teachers who brought me up and taught me everything I know. I am especially thankful to my wife for the enormous help in editing and proofreading this paper. I dedicate this work to my family: my wife Ekaterina, my mother Liubov, and my brother Yaroslav. Most of all, I praise the Holy Trinity for the gifts of music, beauty, and reason.

Glory to the Lord!

# 8 Supplementary materials

## 8.1 Common set-theoretic operations and symbols used in this article

The following is the list of common set-theoretic operations and symbols with descriptions and examples. Those operations are not specific to tuning theory and are the same as in the mathematical literature.

Table 8.1 - Common symbols and operations on sets used in this article.

| **Operation / Symbol** | **Notation** | **Description** |
|---|---|---|
| Size (cardinality) | $\|A\|$ | Number of elements in the set |
| Empty set | $\emptyset$ | A set that has no elements and has a zero size $\|\emptyset\| = 0$. |
| Element of set, Not an element of set | $x \in A$, $x \notin A$ | For example, $x \in A$ is read as "$x$ is an element of the set $A$"; and $x \notin A$ is read as "$x$ is not an element of the set $A$". |
| Subset, Proper Subset | $A \subseteq B$, $A \subset B$ | For two sets $A$ and $B$, $A$ is a subset of $B$ if every element in $A$ is also in $B$. For example, if $A = \{ 1, 2 \}$ and $B = \{ 1, 2, 3 \}$ then $A \subseteq B$. That definition is also true if $A = B$. Sometimes it is necessary to say that $A$ is a subset of $B$ but not equal to $B$. In that case $A$ is called a proper subset of $B$ and is written as $A \subset B$. |
| Superset, Proper Superset | $A \supseteq B$, $A \supset B$ | If $A$ is a subset of $B$, then $B$ is a superset of $A$. For example, $B = \{ 1, 2, 3 \}$ is a superset of $A = \{ 1, 2 \}$. Similarly to the definition of proper subset, when $B \neq A$ we call $B$ a proper |
| | | superset of $A$. For example, a set of whole numbers is a proper superset of natural numbers $\mathbb{Z} \supset \mathbb{N}$. |
| Union | $A \cup B$ | Union of two sets $A$ and $B$ is a set that contains all the elements that are in $A$ or in $B$ or in both $A$ and $B$. Formal definition $A \cup B = \{x \mid x \in A \ or \ x \in B\}$. For example, if $A = \{1, 2, 3\}$ and $B = \{3, 4\}$, then the union $A \cup B = \{ 1, 2, 3, 4 \}$. Note that number 3 is not repeated as by definition, elements in a set are always unique. |
| Intersection | $A \cap B$ | Intersection of two sets $A$ and $B$ is a set that contains all the elements that are shared between $A$ and $B$. Formal definition looks following $A \cap B = \{x \mid x \in A \ and \ x \in B\}$. For example, if $A = \{1, 2, 3\}$ and $B = \{3, 4\}$ then intersection $A \cap B = \{ 3 \}$. |
| Ordered pair | $\langle a, b \rangle$ | The elements in a set are unordered, meaning that $\{a, b\}$ and $\{b, a\}$ is the same set. Sometimes it is helpful to keep the order of elements. Ordered pair $\langle a, b \rangle$ is not the same as $\langle b, a \rangle$. |

## 8.2 Tuning theory specific set theoretic operations and symbols

Most of the labeling of set-theoretic operations used in this article follows the usual convention. However, to perform common tasks within a tuning theory, such as transposition on an interval, required adding some new operations and labels that are not common in the broader literature. The following table lists the tuning theory specific operations used in this article.

Table 8.2 - The list of tuning theory specific operations and symbols used in this article

| | | |
|---|---|---|
| Transposition on an interval (scalar multiplication) | $tF$ or $t \cdot F$ | Defines an operation of transposition of the set $F$ on an interval $t$. The transposition is done by multiplying each frequency in $F$ on $t$. Formal definition: $tF = \{tf \mid f \in F\}$. For example, to transpose a set $F = \{1, 2, 3\}$ up an octave $t = 2$, we can write $2 \cdot F = \{2, 4, 6\}$. This operation can also be called a scalar multiplication. |
| Elementary harmonic sets | $\mathbb{N}_k$ | The symbol $\mathbb{N}$ refers to the set of all natural numbers $\mathbb{N} = \{1, 2, 3, \dots\}$, the index $k$ is used to limit the number of elements in set the following way $\mathbb{N}_k = \{n \mid n \in \mathbb{N}\ n \leq k\}$. For example, $\mathbb{N}_3 = \{1, 2, 3\}$. Such sets are handy to represent harmonic sounds by transposing them on the fundamental frequency (see Formula 3.4). |
| Least common multiple | $\text{lcm}(F)$ | Least common multiple of a set $F$ is the smallest number that is divisible without remainder on every element of the set $F$. For example, $\text{lcm}(\{10, 15\}) = 30$ as $\frac{30}{10} = 3$ and $\frac{30}{15} = 2$. |
| Greatest common divisor | $\mathrm{gcd}(F)$ | Greatest common divisor of a set $F$ is the largest number to which every element of the set $F$ is divisible without remainder. For example, $\mathrm{gcd}(\{10, 15\}) = 5$ as $\frac{10}{5} = 2$ and $\frac{15}{5} = 3$. |
| Note notation | $A4_6$, $C\#5_3$ | This notation is used to indicate the musical note the set refers to and the number of harmonics in that set. For example, the set $A4_6$ stands for a note $A4$ with $6$ harmonics. |
| Chord notation | $A4\text{-dur}_6$ | This notation can be used to indicate the chord. The subscript indicates the number of partials in each note. |
| Pairwise division | $\frac{F_1}{F_2}$ | A set containing a result of the arithmetic division of every element of the set $F_1$ on every element in the set $F_2$. For example, if $F_1 = \{1, 2, 3\}$ and $F_2 = \{5, 7\}$, then $\frac{F_1}{F_2} = \left\{\frac{1}{5}, \frac{2}{5}, \frac{3}{5}, \frac{1}{7}, \frac{2}{7}, \frac{3}{7}\right\}$. |

# 8.3 Frequencies of notes in standard 440Hz tuning

Table 8.3 – Frequency ranges of musical notes

| | |
|---|---|
| **C0** | 15.88 – 16.82 Hz |
| **C#0 / Db0** | 16.82 – 17.82 Hz |
| **D0** | 17.82 – 18.88 Hz |
| **D#0 / Eb0** | 18.89 – 20.01 Hz |
| **E0** | 20.01 – 21.2 Hz |
| **F0** | 21.2 – 22.46 Hz |
| **F#0 / Gb0** | 22.46 – 23.79 Hz |
| **G0** | 23.8 – 25.21 Hz |
| **G#0 / Ab0** | 25.22 – 26.72 Hz |
| **A0** | 26.71 – 28.3 Hz |
| **A#0 / Bb0** | 28.31 – 29.99 Hz |
| **B0** | 29.99 – 31.77 Hz |
| **C1** | 31.76 – 33.65 Hz |
| **C#1 / Db1** | 33.66 – 35.66 Hz |
| **D1** | 35.66 – 37.78 Hz |
| **D#1 / Eb1** | 37.78 – 40.02 Hz |
| **E1** | 40.02 – 42.4 Hz |
| **F1** | 42.4 – 44.92 Hz |
| **F#1 / Gb1** | 44.93 – 47.6 Hz |
| **G1** | 47.6 – 50.43 Hz |
| **G#1 / Ab1** | 50.43 – 53.43 Hz |
| **A1** | 53.43 – 56.61 Hz |
| **A#1 / Bb1** | 56.61 – 59.97 Hz |
| **B1** | 59.98 – 63.54 Hz |
| **C2** | 63.54 – 67.32 Hz |
| **C#2 / Db2** | 67.32 – 71.33 Hz |
| **D2** | 71.32 – 75.57 Hz |
| **D#2 / Eb2** | 75.56 – 80.05 Hz |
| **E2** | 80.06 – 84.82 Hz |
| **F2** | 84.82 – 89.86 Hz |
| **F#2 / Gb2** | 89.86 – 95.21 Hz |
| **G2** | 95.21 – 100.87 Hz |
| **G#2 / Ab2** | 100.87 – 106.87 Hz |
| **A2** | 106.86 – 113.22 Hz |
| **A#2 / Bb2** | 113.22 – 119.95 Hz |
| **B2** | 119.95 – 127.08 Hz |
| **C3** | 127.08 – 134.64 Hz |
| **C#3 / Db3** | 134.64 – 142.65 Hz |
| **D3** | 142.65 – 151.13 Hz |
| **D#3 / Eb3** | 151.13 – 160.11 Hz |
| **E3** | 160.11 – 169.63 Hz |
| **F3** | 169.63 – 179.72 Hz |
| **F#3 / Gb3** | 179.73 – 190.42 Hz |
| **G3** | 190.42 – 201.74 Hz |
| **G#3 / Ab3** | 201.73 – 213.73 Hz |
| **A3** | 213.73 – 226.44 Hz |
| **A#3 / Bb3** | 226.44 – 239.9 Hz |
| **B3** | 239.91 – 254.17 Hz |
| **C4** | 254.18 – 269.29 Hz |
| **C#4 / Db4** | 269.28 – 285.3 Hz |
| **D4** | 285.3 – 302.26 Hz |
| **D#4 / Eb4** | 302.27 – 320.24 Hz |
| **E4** | 320.24 – 339.28 Hz |
| **F4** | 339.28 – 359.46 Hz |
| **F#4 / Gb4** | 359.45 – 380.83 Hz |
| **G4** | 380.84 – 403.48 Hz |
| **G#4 / Ab4** | 403.47 – 427.46 Hz |
| **A4** | 427.47 – 452.89 Hz |
| **A#4 / Bb4** | 452.88 – 479.81 Hz |
| **B4** | 479.82 – 508.35 Hz |
| **C5** | 508.35 – 538.58 Hz |
| **C#5 / Db5** | 538.58 – 570.61 Hz |
| **D5** | 570.6 – 604.53 Hz |
| **D#5 / Eb5** | 604.53 – 640.48 Hz |
| **E5** | 640.49 – 678.57 Hz |
| **F5** | 678.57 – 718.92 Hz |
| **F#5 / Gb5** | 718.92 – 761.67 Hz |
| **G5** | 761.67 – 806.96 Hz |
| **G#5 / Ab5** | 806.96 – 854.94 Hz |
| **A5** | 854.94 – 905.78 Hz |
| **A#5 / Bb5** | 905.78 – 959.64 Hz |
| **B5** | 959.65 – 1016.71 Hz |
| **C6** | 1016.7 – 1077.16 Hz |
| **C#6 / Db6** | 1077.16 – 1141.21 Hz |
| **D6** | 1141.21 – 1209.07 Hz |
| **D#6 / Eb6** | 1209.08 – 1280.97 Hz |
| **E6** | 1280.97 – 1357.14 Hz |
| **F6** | 1357.14 – 1437.84 Hz |
| **F#6 / Gb6** | 1437.84 – 1523.34 Hz |
| **G6** | 1523.34 – 1613.92 Hz |
| **G#6 / Ab6** | 1613.92 – 1709.89 Hz |
| **A6** | 1709.89 – 1811.57 Hz |
| **A#6 / Bb6** | 1811.57 – 1919.29 Hz |
| **B6** | 1919.29 – 2033.41 Hz |
| **C7** | 2033.41 – 2154.32 Hz |
| **C#7 / Db7** | 2154.33 – 2282.43 Hz |
| **D7** | 2282.43 – 2418.15 Hz |
| **D#7 / Eb7** | 2418.16 – 2561.95 Hz |
| **E7** | 2561.94 – 2714.28 Hz |
| **F7** | 2714.29 – 2875.69 Hz |
| **F#7 / Gb7** | 2875.69 – 3046.69 Hz |
| **G7** | 3046.68 – 3227.84 Hz |
| **G#7 / Ab7** | 3227.85 – 3419.79 Hz |
| **A7** | 3419.79 – 3623.14 Hz |
| **A#7 / Bb7** | 3623.14 – 3838.58 Hz |
| **B7** | 3838.59 – 4066.84 Hz |
| **C8** | 4066.84 – 4308.66 Hz |
| **C#8 / Db8** | 4308.66 – 4564.87 Hz |
| **D8** | 4564.87 – 4836.31 Hz |
| **D#8 / Eb8** | 4836.31 – 5123.89 Hz |

## 8.4 Thomae's Function

Thomae's function is a mathematical function that is zero at the irrational numbers and the reciprocal of the denominator at rational numbers:

$$\tau(t) = \begin{cases} \dfrac{1}{q} \;\; if\ t = \dfrac{p}{q} \;\; (t\ is\ rational), with\ p \in \mathbb{Z}\ and\ q \in \mathbb{N}\ coprime \\ 0 \;\; if\ t\ is\ irrational \end{cases} \qquad 8.1$$

Below is the graph of Thomae's function

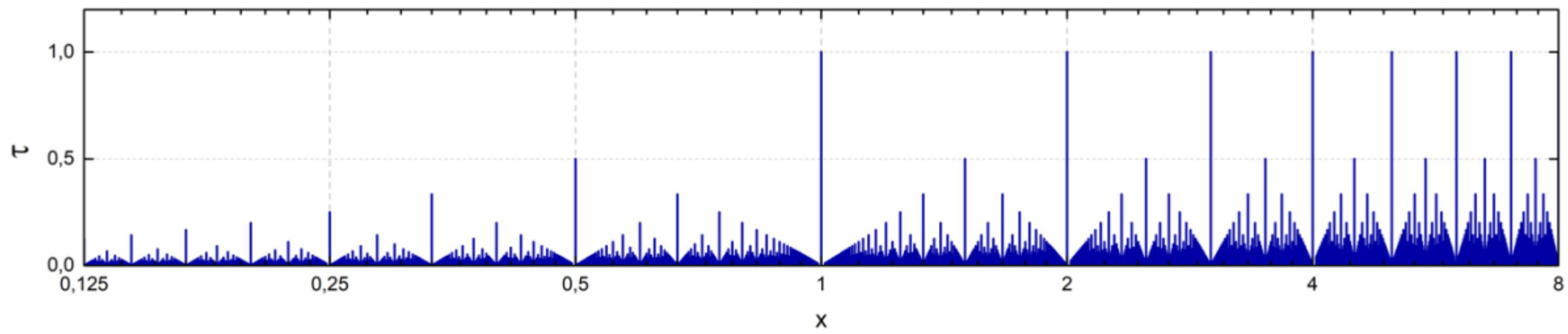

Figure 8.1 - Thomae's function as described by equation 8.1. Note that for $x < 1$ the graph is identical to Figure 5.5.

## 8.5 Proof of total consonance as the inverse of the interval's complexity for a single frequency sound

Here, we prove that the total consonance distribution can be described by Formula 5.9 when the contextual and complementary sets have a single frequency $F = F' = f\mathbb{N}_1$. To demonstrate that, let's use Formula 5.6 to find a total consonance when the complementary set is transposed on an arbitrary interval $t$:

$$\sigma(f\mathbb{N}_1, tf\mathbb{N}_1) = \frac{\alpha(f\mathbb{N}_1, tf\mathbb{N}_1)}{2} + \frac{\chi(f\mathbb{N}_1, tf\mathbb{N}_1)}{2} \tag{8.2}$$

If $t$ is irrational, then both affinity and harmonicity will equal to 0, just like in the Formula 5.9. In the rational case, when $t = \frac{p}{q}$ with $p \in \mathbb{Z}$ and $q \in \mathbb{N}$ coprime, we get the following formula for total consonance:

$$\sigma\left(f\mathbb{N}_1, \frac{p}{q}f\mathbb{N}_1\right) = \frac{\left|f\mathbb{N}_1 \cap \frac{p}{q}f\mathbb{N}_1\right|}{2 \cdot \min\left(|f\mathbb{N}_1|, \left|\frac{p}{q}f\mathbb{N}_1\right|\right)} + \frac{\gcd\left(f\mathbb{N}_1 \cup \frac{p}{q}f\mathbb{N}_1\right) \cdot \left|f\mathbb{N}_1 \cup \frac{p}{q}f\mathbb{N}_1\right|}{2 \cdot \max\left(f\mathbb{N}_1 \cup \frac{p}{q}f\mathbb{N}_1\right)} \tag{8.3}$$

In our case, when both sets in the arguments have only 1 partial, it is always true that $\min\left(|f\mathbb{N}_1|, \left|\frac{p}{q}f\mathbb{N}_1\right|\right) = 1$. Also, the value $f$ can be factored out from $\gcd\left(f\mathbb{N}_1 \cup \frac{p}{q}f\mathbb{N}_1\right) = f \cdot \gcd\left(1, \frac{p}{q}\right)$ and $\max\left(f\mathbb{N}_1 \cup \frac{p}{q}f\mathbb{N}_1\right) = f \cdot \max\left(1, \frac{p}{q}\right)$. Given that, we can simplify our formula of total consonance:

$$\sigma\left(f\mathbb{N}_1, \frac{p}{q}f\mathbb{N}_1\right) = \frac{\left|f\mathbb{N}_1 \cap \frac{p}{q}f\mathbb{N}_1\right|}{2} + \frac{\gcd\left(1, \frac{p}{q}\right) \cdot \left|f\mathbb{N}_1 \cup \frac{p}{q}f\mathbb{N}_1\right|}{2 \cdot \max\left(1, \frac{p}{q}\right)} \tag{8.4}$$

One can notice that $\gcd\left(1, \frac{p}{q}\right) = \frac{1}{q}$ and factoring that additional $q$ in the denominator into the $\max\left(1, \frac{p}{q}\right)$ gives us the following:

$$\sigma\left(f\mathbb{N}_1, \frac{p}{q} f\mathbb{N}_1\right) = \frac{\left|f\mathbb{N}_1 \cap \frac{p}{q} f\mathbb{N}_1\right|}{2} + \frac{\left|f\mathbb{N}_1 \cup \frac{p}{q} f\mathbb{N}_1\right|}{2 \cdot \max\,(p, q)} \qquad 8.5$$

Now let's look at 2 cases when $\frac{p}{q} = 1$ and when $\frac{p}{q} \neq 1$. In the first case, the requirement of $p$ and $q$ being coprime leads to $p = q = 1$ and thus we can write:

$$\frac{p}{q} = 1: \quad \sigma(f\mathbb{N}_1, f\mathbb{N}_1) = \frac{1}{2} + \frac{1}{2 \cdot 1} = 1 \qquad 8.6$$

In the second case, all of the contributions from set sizes have definitive values, and the formula can be simplified:

$$\frac{p}{q} \neq 1: \quad \sigma\left(f\mathbb{N}_1, \frac{p}{q} f\mathbb{N}_1\right) = \frac{0}{2} + \frac{2}{2 \cdot \max(p, q)} = \frac{1}{\max(p, q)} \qquad 8.7$$

However, one can notice that $\frac{1}{\max(p,q)} = 1$ if $p = q = 1$. Thus, the formula for the second case $\frac{p}{q} \neq 1$ can be extended to the case of $\frac{p}{q} = 1$ as it has the same value. Combining all the cases into a single formula, we can write:

$$\sigma(f\mathbb{N}_1, tf\mathbb{N}_1) = \begin{cases} \frac{1}{\max(p, q)} & if\ t = \frac{p}{q}\ is\ rational \\ 0 & if\ t\ is\ irrational \end{cases} \qquad 8.8$$

Which is the same as Formula 5.9, as was to be proved!